\documentclass{jfm}
\usepackage{graphicx}
\usepackage{epstopdf, epsfig}
\usepackage{booktabs}
\usepackage{hyperref}
\usepackage[T1]{fontenc}

\shorttitle{Fourier Triad phase dynamics in 1D Burgers}
\shortauthor{Brendan P.  Murray and Miguel D. Bustamante}

\title{Energy flux enhancement, intermittency and turbulence via Fourier triad phase dynamics in 1D Burgers equation}

\author{Brendan P.  Murray
 \and Miguel D. Bustamante\corresp{\email{miguel.bustamante@ucd.ie}}}

\affiliation{Institute for Discovery, Department of Mathematics and Statistics, University College Dublin, Belfield D4, Ireland}

\begin{document}

\maketitle

\begin{abstract}
We present a theoretical and numerical study of Fourier space triad phase dynamics in one-dimensional stochastically forced Burgers equation at Reynolds number $\mathrm{Re} \approx 2.7 \times 10^4$.
We demonstrate that Fourier triad phases over the inertial range display a collective behaviour characterised by intermittent periods of synchronisation and alignment, reminiscent of Kuramoto model \citep{Kuramoto84} and directly related to collisions of shocks in physical space. These periods of synchronisation favour efficient energy fluxes across the inertial range towards small scales, resulting in strong bursts of dissipation and enhanced coherence of Fourier energy spectrum. The fast time scale of the onset of synchronisation relegates energy dynamics to a passive role: this is further examined using a reduced system with the Fourier amplitudes fixed in time -- a \mbox{phase-only} model. We show that intermittent triad phase dynamics persists without amplitude evolution and we broadly recover many of the characteristics of the full Burgers system. In addition, for both full Burgers and phase-only systems the physical space velocity statistics reveals that triad phase alignment is directly related to the non-Gaussian statistics typically associated with structure-function intermittency in turbulent systems.

\end{abstract}


\setcounter{footnote}{2}
		
Turbulence phenomenology provides equations for the energy budget over spatial scales, based on a statistical balance between the injected energy at large scales, its flux across spatial scales and its subsequent dissipation at small scales. Theoretical and numerical approaches to turbulence share a popular setup in terms of Fourier-space variables that assumes periodic boundary conditions in the spatial coordinates. Seminal results started by Kolmogorov in 1941 (see \cite{uriel1995turbulence}) provide statistical predictions for the energy spectrum: the modulus squared of the complex Fourier amplitudes $E_k = |\hat{u}_k|^2$, while nothing is stated about its conjugate variable, the Fourier phase $\phi_k = \arg (\hat{u}_k)$. A similar situation occurs in wave turbulence theory, which provides coupled evolution equations for the energy spectrum variables only, admitting solutions representing flux of energy-like quantities across spatial scales (see \cite{nazarenko2011wave}). 

Fourier phases are useful variables in Galerkin-truncated discrete models in fluids and nonlinear optics based on triad and/or quartet interactions. There, due to the nonlinear coupling between Fourier energies and the so-called triad or quartet phases (defined as appropriate linear combinations of the Fourier phases), energy transfers can be enhanced by changing the initial phases, in both chaotic and integrable regimes \citep{ADDCraik, kim1997chaotic,  bustamante2009effect, harris2012externally, thompson1991nonlinear, trillo1994nonlinear}. Despite their apparent background role in turbulence, Fourier phases play a direct role in energy fluxes \citep{buzzicotti2016phase} and high-order correlation functions, so their study could shed light on the problem of intermittency. In continuous fluid models described by partial differential equations, the collection of Fourier phases over a range of spatial scales has been shown to be relevant to the dynamics, via amplitude-phase synchronisation mechanism leading to spatiotemporal chaos in channel flows and magnetised Keplerian shear flows \citep{chian2010amplitude, miranda2015off}, and via evolution of phase entropy in cosmological density perturbation solutions \citep{Chiang00}. Motivated by our recent results on triad phase synchronisation in the 2D barotropic vorticity equation and the 1D Burgers equation \citep{bustamante14,buzzicotti2016phase}, our aim is to develop a theory for triad phase dynamics in the context of classical fluid turbulence. This paper provides the first steps in that direction by looking at the behaviour of triad phases in the stochastically-forced 1D Burgers equation
\begin{equation}
\label{eq:burgers}
\frac{\partial u}{\partial t} + \frac{1}{2}\frac{\partial u^2}{\partial x} = \nu\frac{\partial^2 u}{\partial x^2} + f,  \qquad x \in [0,2\pi), \qquad t \in [0,T) \,,
\end{equation}
where $u(x,t) (=u(x+2\pi,t))$ is the scalar velocity field, $\nu$ is a positive parameter (viscosity) and $f(x,t) (=f(x+2\pi,t))$ is the external forcing. 

Burgers equation, see \cite{becreview} for a review, is used as a model for a large range of nonlinear dissipative systems and represents one of the simplest nonlinear partial differential equations known to display a non-trivial scaling of the velocity field correlation functions. Multiscaling is connected to the tendency to create shocks and consequently to increase negative velocity differences $\delta_r v < 0$ and decrease positive ones $\delta_r v > 0$, where $\delta_r v = u(x+r,t) - u(x,t)$. Thus a strongly non-Gaussian probability density function (PDF) of $\delta_r v$ is observed with large tails to the negative side. The presence of strongly localised velocity jumps (shocks) in real space is the fingerprint of Burgers dynamics. We remark that shocks are the only structures in the flow able to dissipate energy in the limit of small viscosity. In other words, the energy flux across scales is absorbed only by a few strongly localised events in real space.

In Section \ref{sec:formulation} we provide the necessary tools. In Section \ref{sec:FPresults} we establish that triad phases over the inertial range synchronise intermittently and we show the contribution of these synchronised events to the energy flux and dissipation. This exploration introduces a Kuramoto-like triad phase order parameter \citep{Kuramoto84} that measures instantaneous synchronisation of triad phases over the inertial range. We will examine its time evolution and split the time averaged statistics of classical quantities into conditional time windows, defined in terms of the level of triad phase synchronisation. This will allow an evaluation of the role of the triad phase dynamics on the intermittent behaviour of the system and on the excursions to the vicinity of fixed points in dynamical systems viewpoint. In Section \ref{sec:FPExtraresults} we demonstrate that these conclusions are robust by extending the range of forcing scales, leading to synchronisation scenarios in the presence of many shocks, and we provide further evidence of the correlation between fast synchronisation events and bursts of energy dissipation. Next, in order to better understand the importance of the role of phase dynamics, in Section \ref{sec:POresults} we define a phase-only Burgers model which removes the time evolution of the Fourier amplitudes while allowing the phases to continue to evolve. Here, we establish that the intermittent phase synchronisation and alignment events seen in the full Burgers PDE persist: notably, the structure-function anomalous exponents and real space non-Gaussian velocity differences are broadly recovered (see also \cite{wilczek2017emergence}).  We show that this phase-only model is quite sensitive to changes in the prescribed energy spectrum scaling, which could explain previous results in fractal Fourier decimated Burgers simulations \citep{buzzicotti2016phase,michelehydro}. We provide a further connection with dynamical systems point of view by demonstrating that the fully-synchronised state is in fact a fixed point of the phase-only model's evolution equations in the unforced case, and show that the stability of this fixed point is sensitive to changes in the prescribed energy spectrum, in a way that is similar to the direct numerical simulation results in the forced case.  In Section \ref{sec:conclusion} we provide concluding remarks and perspectives.

\section{Formulation}
\label{sec:formulation}
\noindent \textbf{Evolution equations.} We introduce the Fourier mode variables $\hat {u}_{k}(t)$, the Fourier space form of the real space solution field $u(x,t)$, where $u(x,t) =\sum_{k \in \mathbb{Z}} e^{i k x} \hat {u}_{k}(t)$ with $x$ the position in real space and $k$ the wavenumbers. The restriction $k\in \mathbb{Z}$ derives from the periodic boundary condition $u(x,t) = u(x+2\pi,t)$. The reality condition $u(x,t) \in \mathbb{R}$ leads to $\hat {u}_{-k}(t) = \hat {u}_{k}^*(t)$, where ${}^*$ denotes complex conjugation. Each mode is indexed by an integer wavenumber $k$.  The dynamical content of mode $k$ is given by its complex valued Fourier mode, $\hat u_k(t)$. Using this representation, the governing PDE can be decomposed into a set of ordinary differential equations (ODEs) which describe the individual time evolution of each of the complex Fourier modes. To delve further into the dynamics we use an amplitude/phase representation: $\hat u_k(t) = a_k(t) e^{i \phi_k(t)}$, where  $a_k(t) = |\hat u_k(t)|$ (mode amplitude) and $\phi_{k}(t) = \arg [\hat{u}_{k}(t)]$ (mode phase).
We examine equation (\ref{eq:burgers}) in Fourier space:
$\frac{\partial \hat {u}_k}{\partial t} = -\frac{ik}{2} \sum_{k_1}\, \hat {u}_{k_1} \hat {u}_{k-k_1} - \nu k^2 \hat {u}_k + \hat {f}_k,
$
 where energy is dissipated via the viscous term and injected by external forcing $\hat {f}_k$. The core of the dynamics is the quadratic non-linear convolution, a term that globally conserves energy by redistributing it amongst Fourier modes via \emph{triad} interactions (groups of 3 modes, as demonstrated in \cite{Kraichnan1967} for 2D Navier Stokes). 
In order to study the role of triad phase dynamics in energy flux mechanism we
derive evolution equations for $a_k$ and $\phi_k$:
\begin{eqnarray}
\hspace{-0.1cm}
\label{eq:ampl_Burg}
\frac{d{a}_k}{dt} &=& \frac{k}{2} \, \sum_{k_1,k_2} a_{k_1} a_{k_2} \, \sin(\varphi_{k_1,k_2}^{k}) \, \delta_{k_1+k_2,k} - \nu k^2 a_{k} + \Re(\hat {f}_k \mathrm{e}^{-i \phi_k})\, , \\
\label{eq:phas_Burg}
\frac{d{\phi}_k}{dt} &=& -\frac{k}{2} \, \sum_{k_1,k_2} \frac{a_{k_1} a_{k_2}}{a_k} \,\cos(\varphi_{k_1,k_2}^{k}) \, \delta_{k_1+k_2,k} +  \frac{1}{a_k} \Im(\hat {f}_k \mathrm{e}^{-i \phi_k}).
\end{eqnarray}

\noindent \textbf{Key degrees of freedom and energy budget.} Upon examination of Eqs.~(\ref{eq:ampl_Burg})--(\ref{eq:phas_Burg}), the key dynamical degrees of freedom \emph{away from the forcing scale} consist of the modes' real amplitudes $a_k(t)$ along with the \emph{triad phases}:  
\begin{equation}
{\varphi}_{k_1,\,k_2}^{k_3}(t) = {\phi}_{k_1}(t) + {\phi}_{k_2}(t) - {\phi}_{k_3}(t),
\end{equation}
where the wavenumbers $k_1$,$k_2$,$k_3$ satisfy a `closed-triad' condition: $k_1 + k_2 = k_3.$ The explicit form of the equations of motion for the triad phases is obtained simply as appropriate linear combinations of equation (\ref{eq:phas_Burg}) for the relevant wavenumbers. 

Let us define the energy spectrum and its rate-of-change budget up to wavenumber $k$:
\begin{equation}
 E_{k}(t) \equiv \hat {u}_{k}\hat {u}_{-k} = a_k^2\,, \qquad \frac{d}{dt}\left(\sum_{k'=1}^k E_{k'}(t)\right) = \mathcal{I}(k,t) - \Pi(k,t) - \mathcal{D}(k,t)\,,
\label{eq:def_spec} 
\end{equation}
where $\mathcal{I}(k,t) \equiv \sum_{k'=1}^k \hat {u}_{k'}^* \hat {f}_{k'} + \mathrm{c.c.\,\,}$ is the energy input rate (forcing for small wavenumbers only), $ \mathcal{D}(k,t) \equiv 2 \nu \sum_{k'=1}^k (k')^2 E_{k'}$ is the dissipation rate and $\Pi(k,t)$ is the flux across wavenumber $k$ towards large wavenumbers.  We derived in \citep{buzzicotti2016phase} an explicit expression for $\Pi(k)$ in terms of the variables $a_k$ and ${\varphi}_{k_1,\,k_2}^{k_3}$:
\begin{equation}
\Pi(k,t) = \sum_{k_1=1}^k \sum_{k_3=k+1}^{\infty} 2 k_1 \, {a}_{k_1} {a}_{k_2} {a}_{k_3} \,  \sin(\varphi_{k_1,k_2}^{k_3}). \\ 
\label{eq:flux_sin}
\end{equation}
We remark here that each term in this equation is due to a single triad interaction between $k_1, k_2$ and $k_3$, such that $ k_1 + k_2 = k_3$. This form of $\Pi(k,t)$ implements explicitly the wavenumber ordering $0 < k_1 < k_3, \quad 0 <  k_2 < k_3$. While this choice of ordering is a simple convention, it is a useful one because the triad phases $\varphi_{k_1,k_2}^{k_3}$, \emph{with that ordering}, are found to take values preferentially near $\pi/2 \pmod {2\pi}$. 

As discussed in detail in \cite{buzzicotti2016phase}, equation (\ref{eq:flux_sin}) shows a clear link between the strong probability of triad phases to take a value close to $\pi/2 \pmod{ 2\pi}$ and the efficiency of the energy flux towards small scales. Such a PDF of ${\varphi}_{k_1,\,k_2}^{k_3}$ (see figure \ref{fig:event_approach}, bottom left) is explained by examining the contribution from a given triad (each term in equation (\ref{eq:flux_sin})) to the energy flux towards small scales.
As $2 \, k_1 \, a_{k_1}a_{k_2}a_{k_3} >0$, at any given time a positive contribution to the energy flux is maximised if the triad phase takes the value $\pi/2 \pmod{ 2\pi}$. In Section \ref{sec:FPresults} we show that this triad phase alignment to $\pi/2 \pmod{ 2\pi}$ occurs for all triads in the so-called ``inertial-range'' of scales, which is characterised by a power-law spectrum  $a_k \propto k^{-1}$ (see \cite{becreview}). 

Finally, we define the total dissipation $\epsilon$ and total energy as
\begin{equation}
\epsilon(t) \equiv \mathcal{D}(\infty,t) = 2 \nu \sum_{k=1}^{\infty} k^2 E_k \quad , \quad E(t) = \sum_{k=1}^{\infty}E_{k}.\\ 
\label{eq:def_tot}
\end{equation}

\noindent \textbf{Synchronisation and alignment.} To gain a quantitative measure of the triad phase synchronisation and alignment in time we define \citep{Kuramoto84, Strogatz00} the (complex-valued) Kuramoto order parameter for the triad phases:
\begin{equation}
R \, \exp({i \, \Phi}) = \left< \exp({i \, {\varphi}_{k_1,k_2}^{k_1+k_2}}) \right>_{k_1,k_2}\,, 
\label{eq:phase_order}
\end{equation}
where the average is over triads. Here $R(t) \in [0,1]$ measures the coherence of the triad phase population, and $\Phi(t) \in [0,2\pi)$ is a proxy for the average triad phase. A \emph{synchronised} state of the triad phases is signalled when $R(t) = 1$: the corresponding $\Phi(t)$ denotes the value of the triad phases. \emph{Alignment} occurs when this value is equal to $\pi/2$, which leads to maximised flux contributions in equation (\ref{eq:flux_sin}). Correspondingly, $ R(t) \lesssim 1$ indicates that not all triad phase values are the same but there is still a preference for clustering of the values. In contrast, complete disorder - a uniform probability for the triad phases to take a value in $[0,2\pi)$ - would give $R(t) \approx 0$. By studying the time evolution of $R(t)$ and $\Phi(t)$ for triads in the inertial range we can study how triad synchronisation and alignment directly affect the global characteristics of the system.  \\

\noindent \textbf{Intermittency Statistics.} To establish a link between Fourier-space dynamics and classical intermittency we need to analyse the physical space velocity field statistics, particularly the multi-point correlation functions (structure functions):
\begin{equation}
S^{p}(r) = \langle   \, (\delta_r v)^{p} \, \rangle_{x,t} \qquad , \qquad S_{\mathrm{abs}}^{p}(r) = \left< \, |\delta_r v|^{p} \, \right>_{x,t},\\ 
\label{eq:struct_funcs}
\end{equation}
where $\delta_r v = u(x+r) - u(x)$ and $\left < \bullet \right >_{x,t}$ denotes the average over both space and time. In the inertial range the structure functions follow a power-law in $r$ and we define the scaling exponents $\zeta_p$ and $\zeta_p^{\mathrm{abs}}$ via $S_{\mathrm{abs}}^{p}(r) \sim r^{\, \zeta_p}$ and  $S_{\mathrm{abs}}^{p}(r) \sim r^{\, \zeta_p^{\mathrm{abs}}}$ respectively.  Anomalous exponents ($\zeta_p < p$) are a signature of intermittency. The role of triad phase synchronisation can be studied by computing the structure functions restricted to strong synchronisation events and comparing the obtained exponents with the usual ones; more generally, obtaining the probability density function of velocity increments restricted to synchronisation events of various degrees (measured by the Kuramoto parameter $R(t)$).

\section{Numerical Results for Large Scale Forcing}
\label{sec:FPresults}
We present a numerical study of Burgers triad phase dynamics with Gaussian white-in-time forcing acting only at large spatial scales, with forced wavenumbers $k_f \in [1,3]$. The numerical simulation uses a pseudospectral, $4^{\mathrm{th}}$-order Adams-Bashforth time-stepping scheme, with the number of collocation points $N = 2^{18}$, $k_{\max} = N/3$ and a time step $\delta t \sim 10^{-6}$. We take a viscosity value $\nu = 2 \times 10^{-4}$, leading to average total energy  $\overline{E} \equiv \left< E(t) \right>_t \approx 5.25$ and average energy dissipation rate $\overline{\epsilon} \equiv \langle \epsilon(t) \rangle_t \approx 8.08$. The Kolmogorov length and time microscales are  $\eta \equiv (\nu^{3}/\overline{\epsilon})^{1/4} \approx 0.001$ and $\tau_\eta \equiv (\nu/\overline{\epsilon})^{1/2} \approx 0.005$, and the Reynolds number is $Re \equiv ((2\pi/k_f)/\eta)^{4/3} \approx 2.7\times10^{4}.$ Statistically stationary  variables (Fig.~\ref{fig:vel_incFP}) were averaged over simulation runtime of $6 \times 10^{5} \, \tau_\eta$. 

\begin{figure}
  \vspace{-2mm}
  \centerline{\resizebox{1.20\textwidth}{!}{\includegraphics{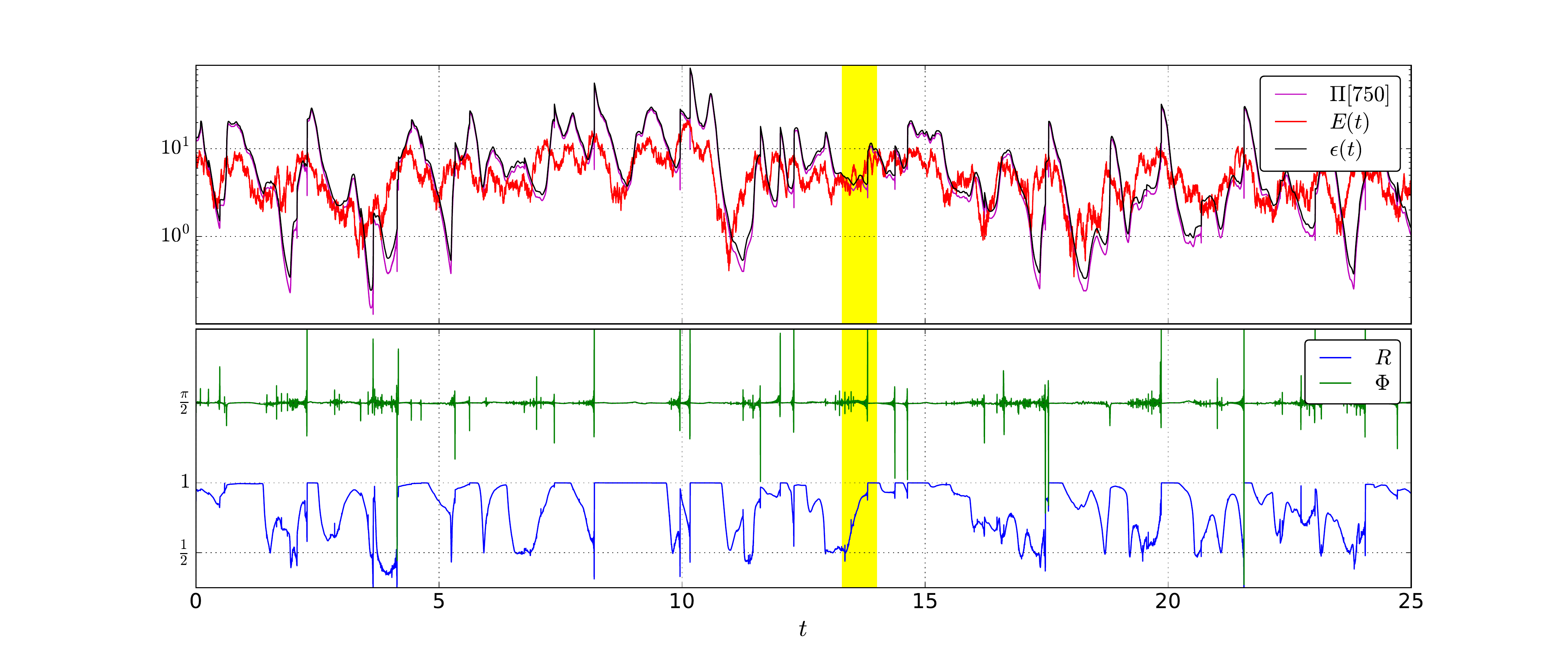}}}
  \vspace{-2mm}
	\caption{Subset of time evolution of  global system quantities, for the simulation under large-spatial-scale stochastic forcing ($k_f \in [1,3]$). Highlighted is the time range $[13.3 \, ,14.0]$, where the time approach to a synchronisation event will be studied in figure \ref{fig:event_approach}. \textbf{Top}: Energy flux $\Pi[k=750](t)$\textit{(magenta)} at the high-wavenumber end of the inertial range. Total energy of the system $E(t)$\textit{(red)} and total energy dissipation rate $\epsilon(t)$\textit{(black)}. \textbf{Bottom}: Triad phase order parameter $R (t)$ \textit{(blue)} and $\Phi(t)$\textit{(green)} for triad phases in the inertial range. 
	\label{fig:tseries}
}
\end{figure}
\begin{figure}
  \centerline{\resizebox{1.05\textwidth}{!}{\includegraphics{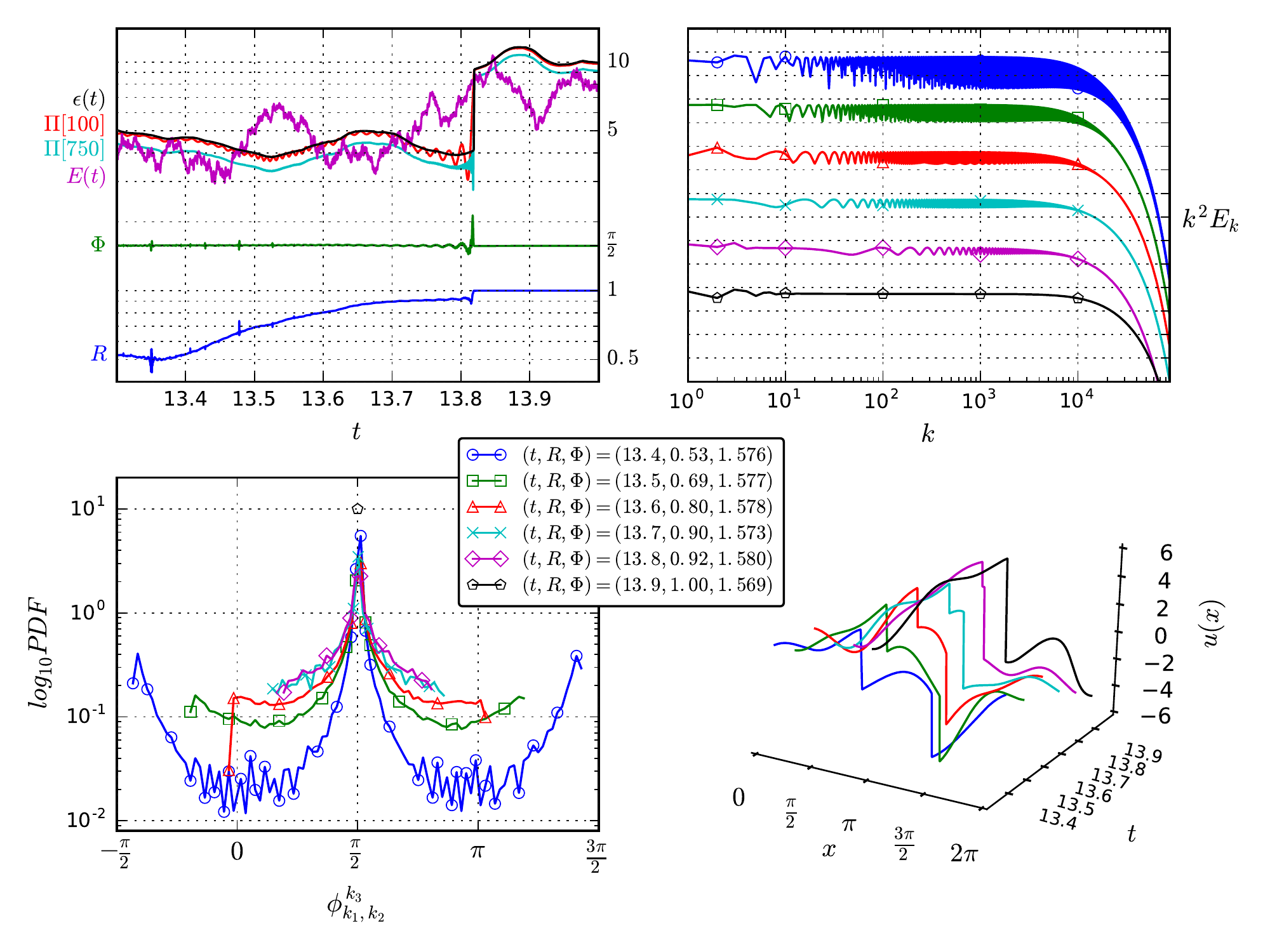}}}
  \caption{Analysis of system dynamics on approach to a synchronisation event ($R\rightarrow1$), for the simulation under large-spatial-scale stochastic forcing ($k_f \in [1,3]$). \textbf{Top Left}: Zoomed time-series of highlighted region in figure \ref{fig:tseries} showing the sharp jump in dissipation and energy flux when triad phases become synchronised. \textbf{Top Right}: $k^2$-scaled energy spectra $E_k$ at six snapshots (shifted for visualisation purposes). \textbf{Bottom Left}: Snapshots of PDF over all inertial-range triad phases ${\varphi}_{k_1,\,k_2}^{k_3}$. \textbf{Bottom Right}: Real-space solution snapshots $u(x,t)$.}
\label{fig:event_approach}
\end{figure}
\begin{figure}
\vspace{2mm}
\resizebox{0.50\textwidth}{!}{
  \includegraphics{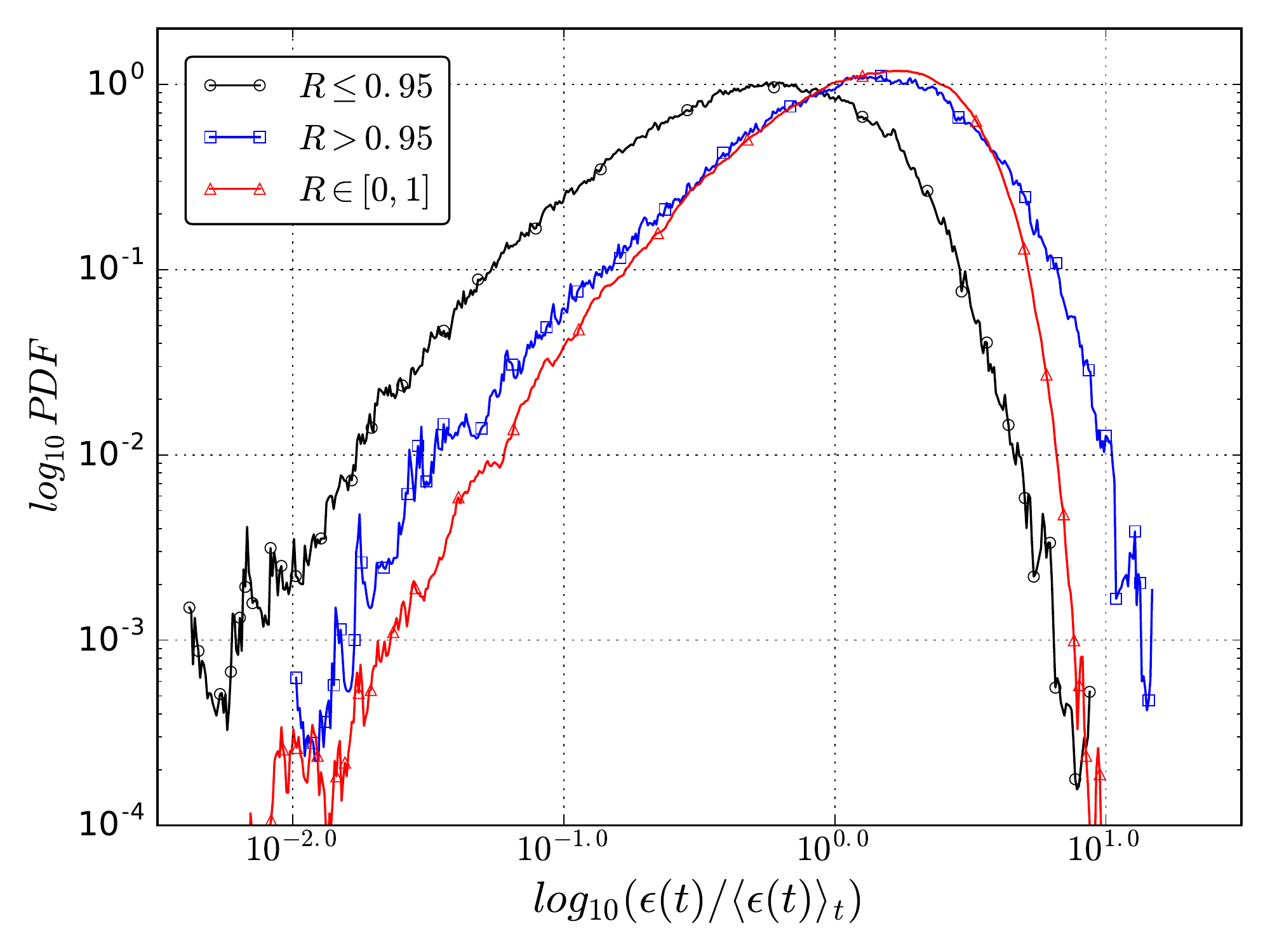}}
\resizebox{0.50\textwidth}{!}{
  \includegraphics{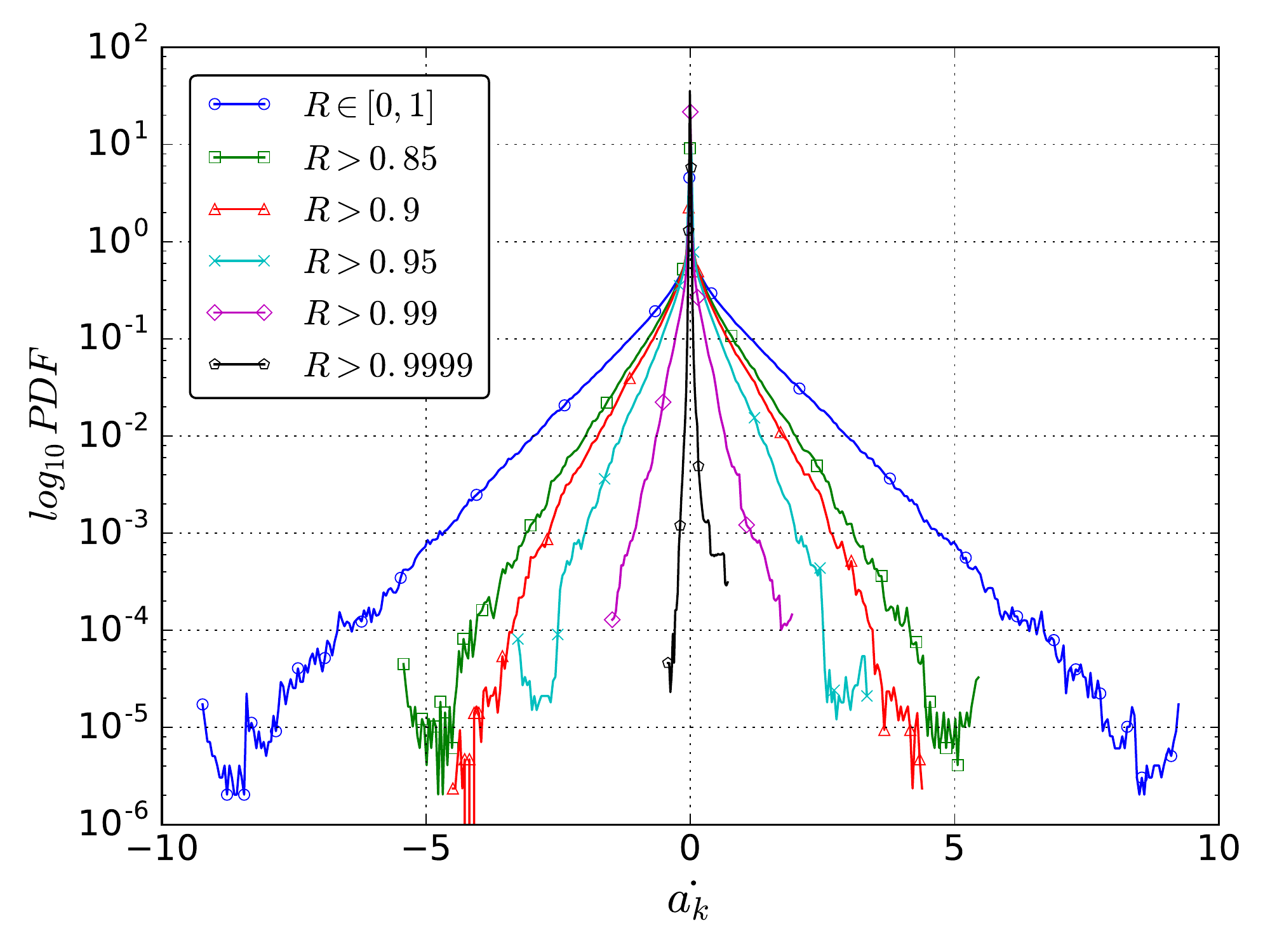}}
\resizebox{0.50\textwidth}{!}{
  \includegraphics{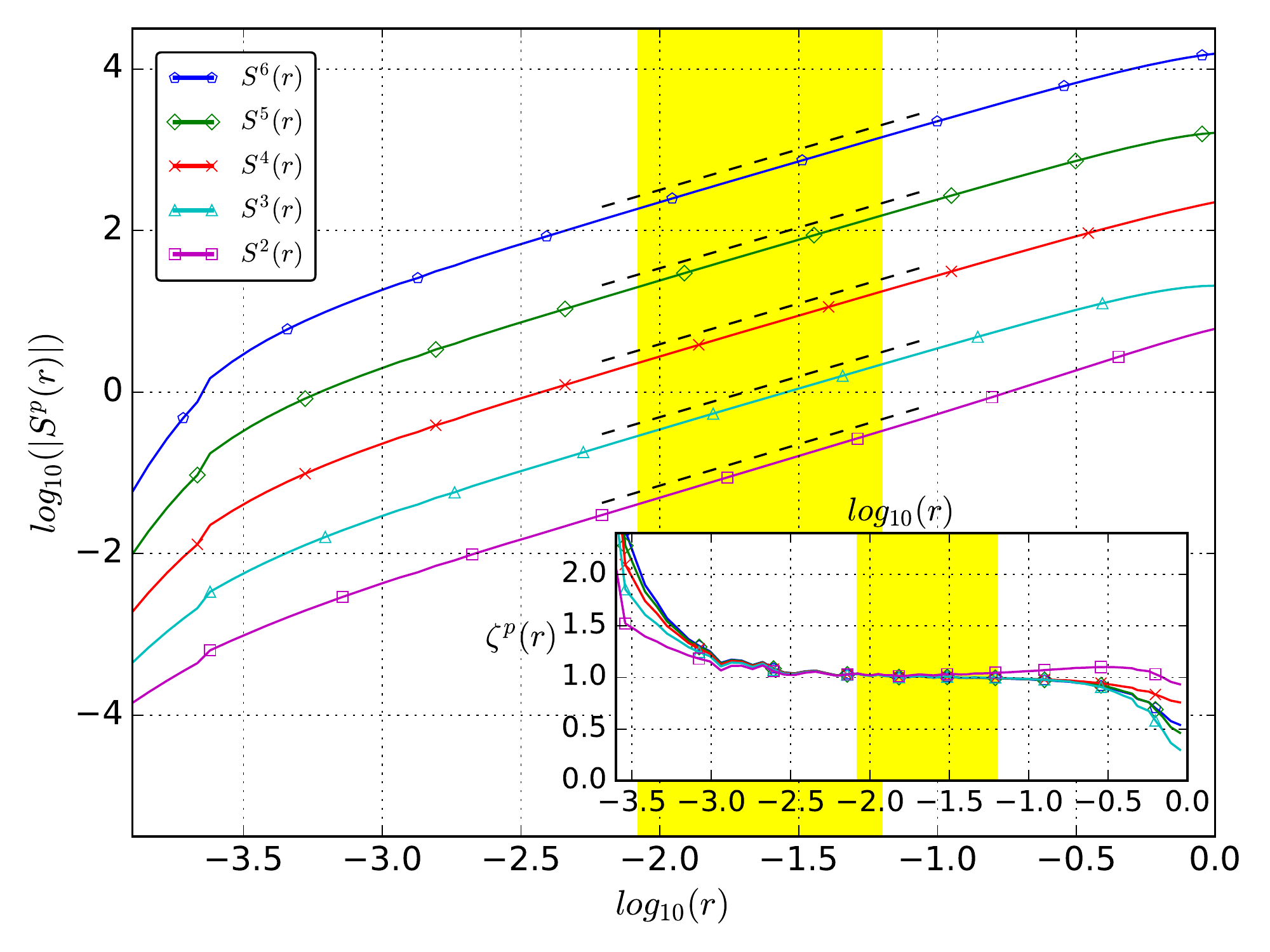}}
\resizebox{0.50\textwidth}{!}{
  \includegraphics{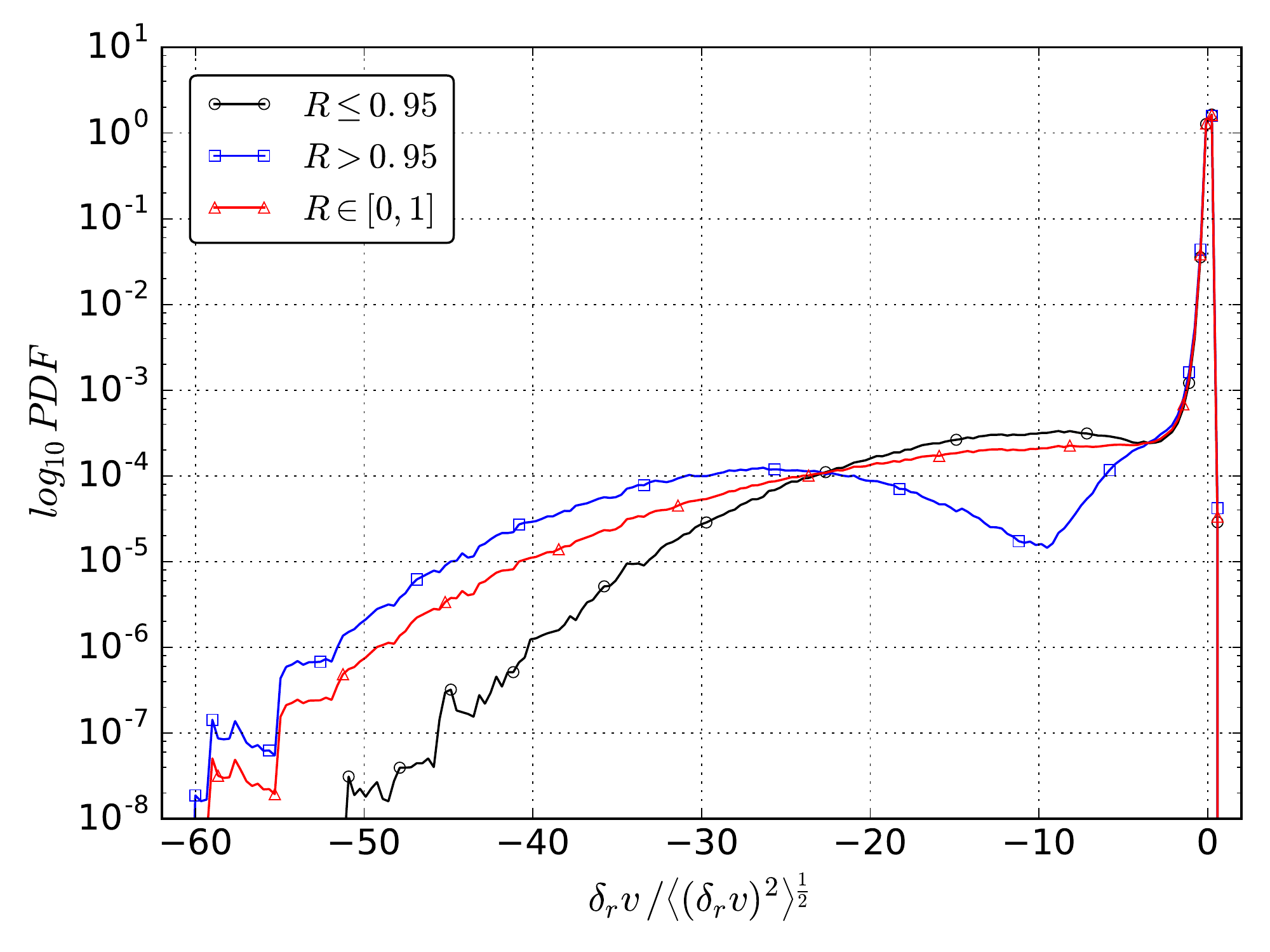}}
\caption{ Statistics over full statistically stationary state, for the simulation under large-spatial-scale stochastic forcing ($k_f \in [1,3]$). \textbf{Top Left}: Conditional time PDF of  total dissipation $\epsilon(t)$ based on value of phase order parameter $R$. \textbf{Top Right}: PDF of $\dot{a}_k$ for a range of conditional time periods. \textbf{Bottom Left}: Structure functions $S^{p}(r)$ as a function of $r$ and (inset)  associated local slopes $\zeta_{p}(r)$. \textbf{Bottom Right}: Conditional time PDFs of the normalised velocity increments $\delta_{r} v \, / \left< (\delta_{r} v)^2\right>^{{1}/{2}}$ in the inertial range at a scale $k=450$ or $r=0.013$. }
\label{fig:vel_incFP}
\end{figure}
Figure \ref{fig:tseries} shows the evolution over a time interval of $5 \times 10^3 \, \tau_\eta$. We observe intermittent in time events of synchronisation of triad phases, signalled by jumps of the Kuramoto order parameter $R(t)$ to values close to $1$ (we consider $R > 0.95$ as synchronised); the synchronisation persists for some time after each event, but this duration is irregular (another signature of intermittency). For each of these events, a clear pattern is found in the energy flux at inertial-range scales $\Pi(750,t), \Pi(100,t)$ (see also figure \ref{fig:event_approach}, top left panel): an early oscillation and then a jump to large values. A similar alignment is observed in the Kuramoto phase $\Phi(t)$, with value close to $\pi/2$ at these events.  While remarkable jumps are seen in the total dissipation $\epsilon(t)$ (Eq.~\ref{eq:def_tot}) at these events, no early oscillation is detected in this variable. Notice also that the classical stationary energy-budget identity $\langle\Pi(100,t)\rangle_t = \langle\epsilon(t)\rangle_t$ [\cite{uriel1995turbulence}] is in fact nearly satisfied at all times, except during the pre-synchronisation oscillations. The total system's energy $E(t)$ (Eq.~\ref{eq:def_tot}) 
does not show any special behaviour preceding the synchronisation events.

Figure \ref{fig:event_approach} shows statistics during a close-up of one such synchronisation event in the time region  highlighted in figure \ref{fig:tseries}. Here we show 6 snapshots in time as the triad phase order parameter $R \rightarrow 1$ with the legend showing the times and phase order parameters $(R,\Phi)$ at each snapshot. The top left panel shows again the time evolution of global system quantities and shows the distinct time at which the system changes from a non-synchronised to synchronised state. The top right panel shows the $k^2$ rescaled spectrum at each of these snapshots (each snapshot has been shifted for visualisation purposes). Upon approach to $R\approx 1$ (top to bottom in plot) we see a clear reduction in the energy fluctuations across wavenumbers, with the final snapshot in the synchronised region showing a uniform $k^{-2}$ scaling in the inertial range. This reduction in amplitudes $a_k$ is further quantified in the top right panel of figure \ref{fig:vel_incFP} and will be discussed in more detail later. The bottom left panel shows snapshots of the PDF of the triad phase variable ${\varphi}_{k_1,\,k_2}^{k_3}$, over all triads in the inertial range $100 < k < 750$. At all snapshots the triad phase preferentially takes values near $\pi/2  \pmod {2\pi}$ but as the synchronisation event is approached the distribution narrows until, at the synchronisation time, all triad phase values are extremely close to $\pi/2  \pmod {2\pi}$, maximising the direct energy flux towards small scales: all terms contribute in a positive and maximal way to the flux in Eq.~(\ref{eq:flux_sin}). The bottom right panel shows the real space solution $u(x,t)$ at each snapshot. From front to back we see the real space signature of the event: two shocks merging. This explains the sharp jump in dissipation seen at each synchronisation event as a large single shock is created. To quantify the jump in dissipation we provide the conditional-time dissipation PDF during synchronisation events over the full simulations of the statistically stationary state in figure \ref{fig:vel_incFP}, top left panel. It is evident that these synchronised events correspond to significantly higher dissipation than non-synchronised ones (by a factor 3), with the highest dissipation events exclusively restricted to these $R \approx 1$ time intervals.

We provide evidence that synchronisation events are correlated with excursions near fixed points in the dynamical systems point of view. A fixed point is signalled by the simultaneous zeroes of $\dot{a}_k$ and $\dot{\varphi}_{k_1 k_2}^{k_3}$, notably across the energetically active inertial range. While equation (\ref{eq:phas_Burg}) implies that triad-phase time derivatives are drastically reduced as the order parameters $(R, \Phi)$ tend to $(1,\pi/2)$,  equation (\ref{eq:ampl_Burg}) does not trivially imply a reduction of the amplitude time derivatives during synchronisation events. Figure \ref{fig:vel_incFP}, top right panel, shows the conditional-time PDF of mode-amplitude time derivatives $\dot{a}_k$ over the inertial range, restricted to events with several levels of synchronisation. The higher the synchronisation($R$), the closer the $\dot{a}_k$  distribution clusters about zero values. Such near absolute synchronisation is unlikely, however we note that the amount of time the system spends in the $R>0.9999$ region is approximately 15\% of the full simulation time.

Finally, we establish the role of synchronisation events $R>0.95$ on the real-space intermittency statistics by examining PDF of velocity increments $\delta_{r} v = u(x+r,t)-u(x,t)$ in the full statistically stationary simulations. The structure functions $S^p(r) \propto r^{\zeta_p}$ and associated local slopes are shown in figure \ref{fig:vel_incFP}, bottom left panel. Using the highlighted region of flat local slopes to fit ($0.008 \leq r \leq 0.063$ which corresponds to our inertial range of $100 \leq k \leq 750$), we recover the usual \citep{michelehydro} scaling exponents (table of fit values and related figures can be found in the Ancillary Material). The bottom right panel shows that the restriction of the PDF to the time periods where $R>0.95$ results in events of larger velocity increments with a significantly higher probability, which correspond to large dissipative shocks and in turn provide important contributions to the structure functions. Fit values for scaling exponents $\zeta_p$ and $\zeta_p^{\mathrm{abs}}$ show that the exponents are significantly closer to the desired value of $1$ for the $R>0.95$ conditional time regions. 

\section{Numerical Results with Additional range of Forcing}
\label{sec:FPExtraresults}
We present here a further numerical study of Burgers triad phase dynamics for a wider scale of Gaussian white-in-time forcing on wavenumbers $k_f \in [1,12]$ rather than the earlier $k_f \in [1,3]$. The other numerical parameters are identical to Section \ref{sec:FPresults} with the exception of the viscosity value which is now $\nu = 3 \times 10^{-4}$, leading to average total energy  $\overline{E} \equiv \left< E(t) \right>_t \approx 1.58$ and average energy dissipation rate $\overline{\epsilon} \equiv \langle \epsilon(t) \rangle_t \approx 1.975$. The Kolmogorov length and time microscales are  $\eta \equiv (\nu^{3}/\overline{\epsilon})^{1/4} \approx 0.002$ and $\tau_\eta \equiv (\nu/\overline{\epsilon})^{1/2} \approx 0.012$, and the Reynolds number is $Re \equiv ((2\pi/k_f)/\eta)^{4/3} \approx 2.07\times10^{3}.$ Statistically stationary  variables (Fig.~\ref{fig:FPJointPDF}) were averaged over simulation runtime of $1 \times 10^{5} \, \tau_\eta$. 

\begin{figure}
  \centerline{\resizebox{1.0\textwidth}{!}{\includegraphics{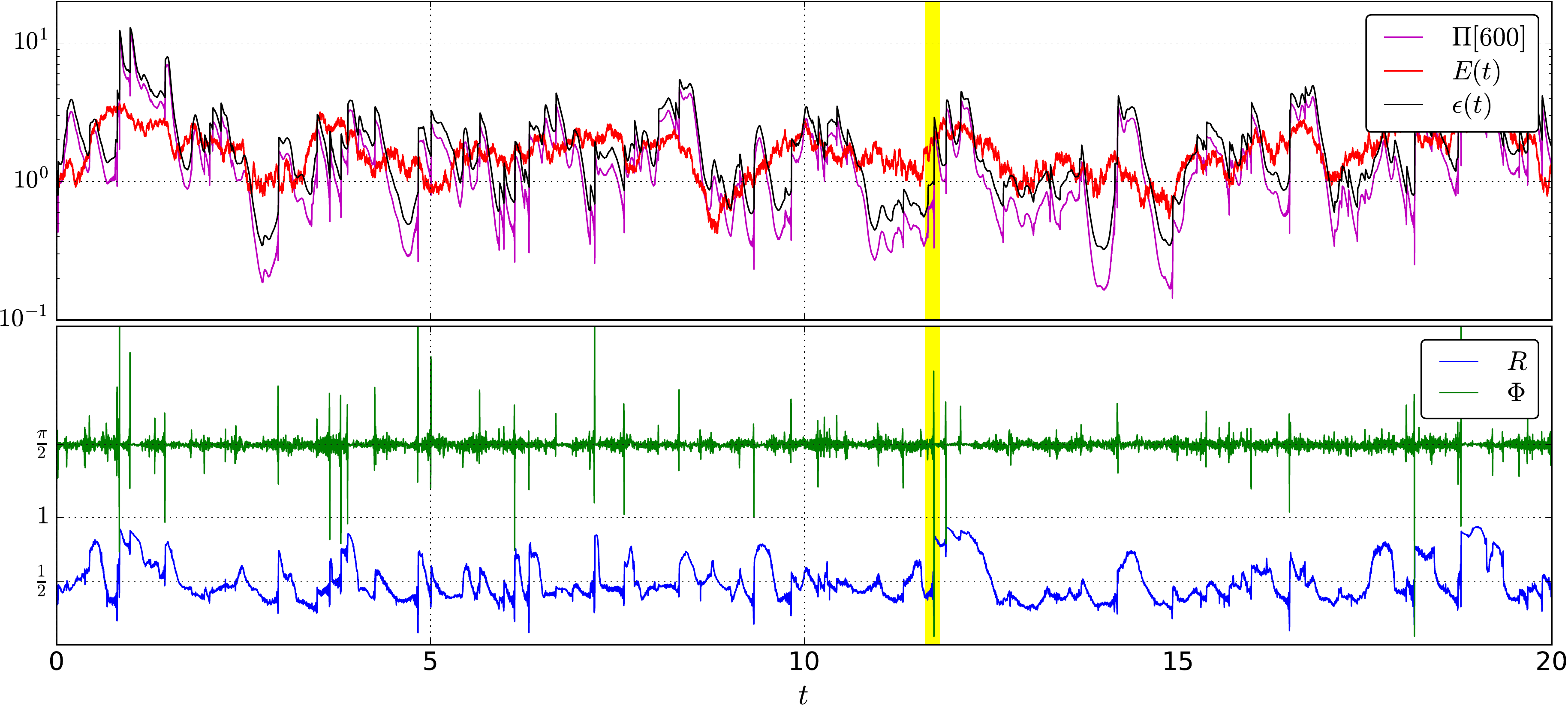}}}
	\caption{Subset of time evolution of global system quantities, for the simulation under extended-spatial-scale stochastic forcing ($k_f \in [1,12]$). Highlighted is the time range $[11.6 \, ,11.85]$, where the time approach to a jump in synchronisation event will be studied in figure \ref{fig:event_approachExtra}. \textbf{Top}: Energy flux $\Pi[k=150](t)$\textit{(red)}, $\Pi[k=600](t)$\textit{(cyan)} at each end of the inertial range. Total energy of the system $E(t)$\textit{(magenta)} and total energy dissipation rate $\epsilon(t)$\textit{(black)}. \textbf{Bottom}: Triad phase order parameter $R (t)$ \textit{(blue)} for triad phases in the inertial range. 
	\label{fig:tseriesExtra}
}
\end{figure}
\begin{figure}
  \centerline{\resizebox{1.0\textwidth}{!}{\includegraphics{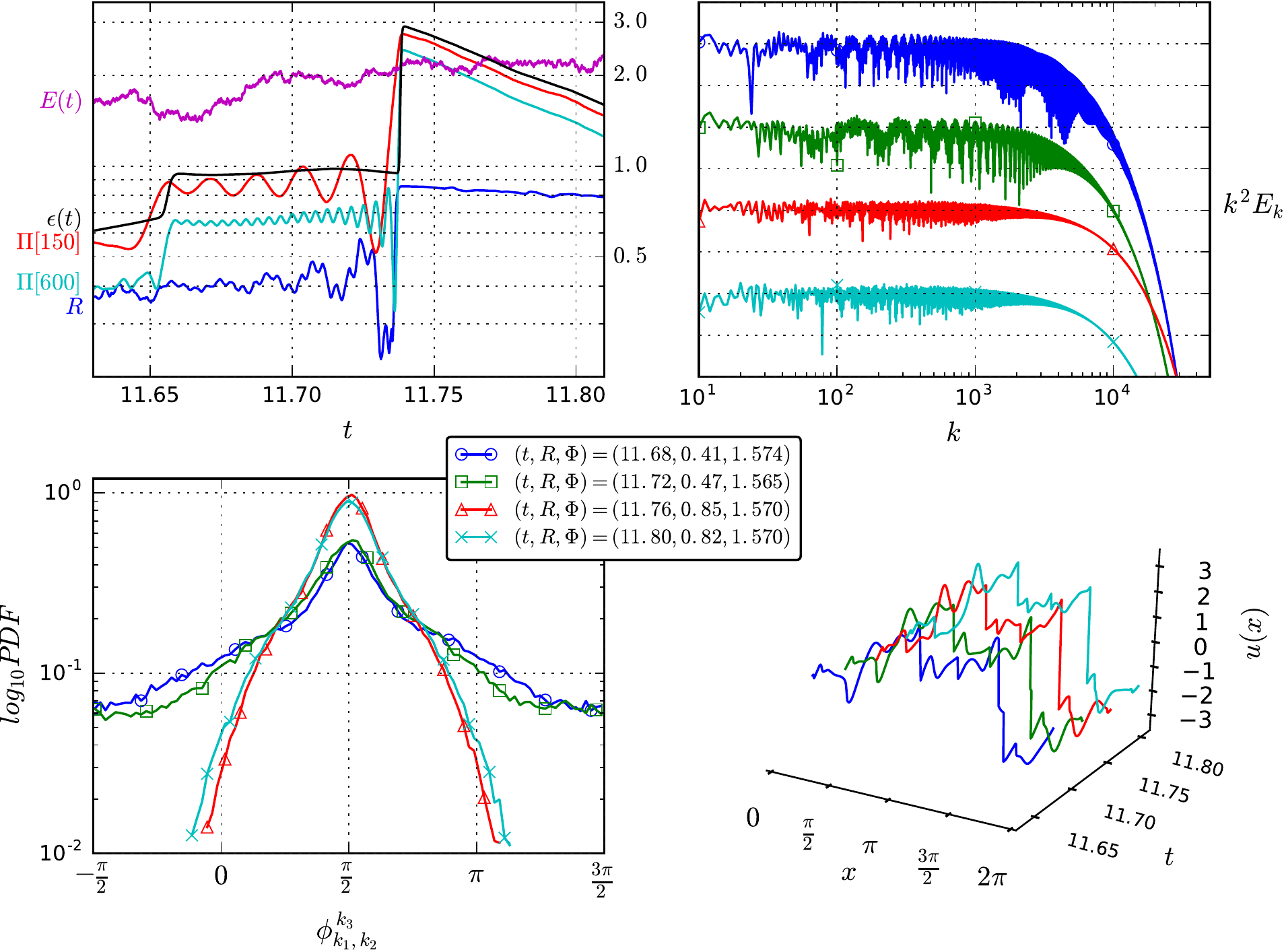}}}
  \caption{Analysis of system dynamics on approach to a partial synchronisation event ($R \rightarrow 0.85$), for the simulation under extended-spatial-scale stochastic forcing ($k_f \in [1,12]$). \textbf{Top Left}: Zoomed time-series of highlighted region in figure \ref{fig:tseriesExtra} showing the sharp jump in dissipation and energy flux when the triad phase synchronisation parameter $R$ increases. \textbf{Top Right}: $k^2$-scaled energy spectra $E_k$ at four snapshots (shifted for visualisation purposes). \textbf{Bottom Left}: Snapshots of PDF over all inertial-range triad phases ${\varphi}_{k_1,\,k_2}^{k_3}$. \textbf{Bottom Right}: Real-space solution snapshots $u(x,t)$.}
\label{fig:event_approachExtra}
\end{figure}

\begin{figure}
  \centerline{\resizebox{0.85\textwidth}{!}{\includegraphics{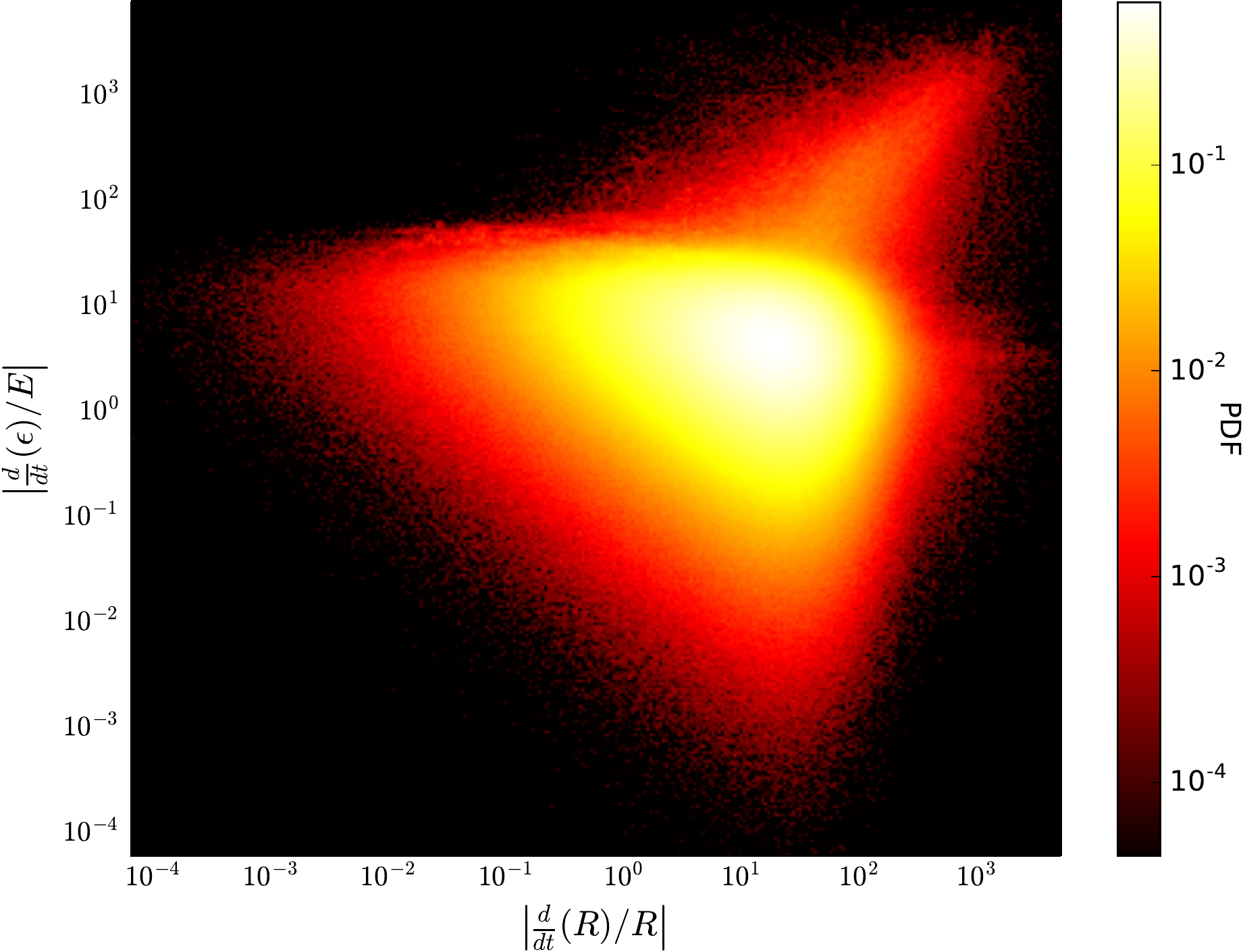}}}
  \caption{Joint PDF, over all times $t$ of our statistically-stationary simulation with stochastic forcing over wavenumbers $k_f \in [1,12]$, between the normalised time derivative of $R$ $(\left| \frac{d}{dt}(R) / R \right|)$ at time $t$ and the normalised time derivative of total system dissipation $ \left| \frac{d}{dt}(\epsilon) / E \right|$ at the slightly later time $t+0.03 \tau_\eta$, where $\tau_\eta$ is the Kolmogorov time microscale. }
\label{fig:FPJointPDF}
\end{figure}

Figure \ref{fig:tseriesExtra} shows the evolution over a time interval of $1 \times 10^3 \, \tau_\eta$. We observe intermittent-in-time events of rapid partial synchronisation of triad phases, signalled by jumps of the Kuramoto order parameter $R(t)$; As in the case of the results presented in Section \ref{sec:FPresults} the synchronisation persists for some time after each event and again this duration is irregular. For each of these events, a clear pattern is found in the energy flux at inertial-range scales $\Pi(600,t), \Pi(150,t)$ (see also figure \ref{fig:event_approachExtra}, top left panel): an early oscillation and then a rapid jump. A similar alignment is observed in the Kuramoto phase $\Phi(t)$, with value close to $\pi/2$ at these events. Similarly to the earlier results, jumps are seen in the total dissipation $\epsilon(t)$ (Eq.~\ref{eq:def_tot}) at these events, confirming that the link between dissipation bursts and the degree of triad phase synchronisation is apparent also for scenarios where many real-space shock structures are present. 

Figure \ref{fig:event_approachExtra} shows statistics during a close-up of one such synchronisation event in the time region  highlighted in figure \ref{fig:tseriesExtra}. Here we show four snapshots in time as the triad phase order parameter $R \rightarrow 0.85$ with the legend showing the times and phase order parameters $(R,\Phi)$ at each snapshot. The top left panel shows again the time evolution of global system quantities and shows the distinct time at which the system changes from a non-synchronised to synchronised state. The top right panel shows the $k^2$ rescaled spectrum at each of these snapshots (each snapshot has been shifted for visualisation purposes). Upon approach to the jump in synchronisation (top to bottom in plot) we see a clear reduction in the energy fluctuations across wavenumbers in both the inertial and dissipative range. The remaining fluctuations are due to the presence of other shocks during the whole process. The bottom left panel shows snapshots of the PDF of the triad phase variable ${\varphi}_{k_1,\,k_2}^{k_3}$, over all triads in the inertial range $150 < k < 600$. At all snapshots the triad phase preferentially takes values near $\pi/2  \pmod {2\pi}$ but as the synchronisation event is approached the distribution narrows as the triad phase values are clustered around $\pi/2  \pmod {2\pi}$, maximising the direct energy flux towards small scales: the vast majority of terms in Eq.~(\ref{eq:flux_sin}) contribute in a positive and maximal way to the flux. The bottom right panel shows the real space solution $u(x,t)$ at each snapshot. From front to back we see the real space signature of the event: two shocks merging. This is  similar to the results presented in Section \ref{sec:FPresults}, however here we have the presence of many other shocks in the system throughout this event which results in an imperfect synchronisation. 

Unlike the earlier results in figure \ref{fig:vel_incFP} which examine cases of full triad phase alignment which persisted after the event, in these partial synchronisation events we have a more complex criterion for capturing the statistical link between the increase in triad phase synchronisation and the change in total dissipation. Rather than monitoring $R$ directly, in figure \ref{fig:FPJointPDF} we show a joint PDF between the normalised time derivative of $R$, $\left| \frac{d}{dt}(R) / R \right|$, and the normalised time derivative of total system dissipation $ \left| \frac{d}{dt}(\epsilon) / E \right|$. Here we also need to take account of the \textit{time lag} between the moment when the triad phase synchronisation occurs, the subsequent burst in inertial-range energy flux and the dissipation that follows. For the numerical parameters of this particular study we estimated this time lag to be $\approx 0.03 \tau_\eta$ and this time lag is applied to the pair value that makes up figure \ref{fig:FPJointPDF}. The relationship can be clearly seen between the rate of change of the total energy dissipation and the rapid jumps in phase order parameter $R$. Similar to the full synchronisation results, we see that the dynamics of the phase order parameter are key in all highly dissipative energy bursts in the system.  

\section{Phase-Only Model}
\label{sec:POresults}

The previous results demonstrate that the events of triad phase synchronisation and alignment are common (albeit intermittent) in stochastically-forced simulations. Moreover, these events produce dramatic energy flux and dissipation enhancements. In this Section we produce two interrelated studies: first, we quantify how much of the alignment behaviour observed is driven by the forced phase dynamics only. Second, we provide a quantitative stability analysis of the fixed point (in dynamical systems point of view) corresponding to the fully-aligned, fully-synchronised state.  

\subsection{Stochastically forced phase dynamics}
We approach the question by ``freezing'' the amplitudes $a_k$ to take prescribed values. This means we drop Eq.~(\ref{eq:ampl_Burg}) and keep Eq.~(\ref{eq:phas_Burg}):
\begin{eqnarray}
\hspace{-0.1cm}
\label{eq:phas_only}
\dot{\phi}_k &=& -\frac{k}{2} \, \sum_{k_1,k_2} \frac{a_{k_1} a_{k_2}}{a_k} \,\cos(\varphi_{k_1,k_2}^{k}) \, \delta_{k_1+k_2,k}  + \frac{1}{a_k} \Im(\hat {f}_k \mathrm{e}^{-i \phi_k}).
\end{eqnarray}
We prescribe the amplitudes as the time-averaged spectrum obtained from the full PDE simulation, thereby removing the dependence on spectrum fluctuations evidenced in figure \ref{fig:tseries}. Initial phases are taken from the end snapshot of full PDE simulations, thereby obtaining a realistic initial triad phase distribution for the phase-only dynamical system (\ref{eq:phas_only}). Stochastic forcing at large scales (wavenumbers 1 to 3) is again used as in Section \ref{sec:FPresults} but the forced mode amplitudes are restored to their prescribed values before moving at each time step. In this way the individual phases will evolve effectively as in equation (\ref{eq:phas_only}).  

\begin{figure}
\resizebox{0.50\textwidth}{!}{
  \includegraphics{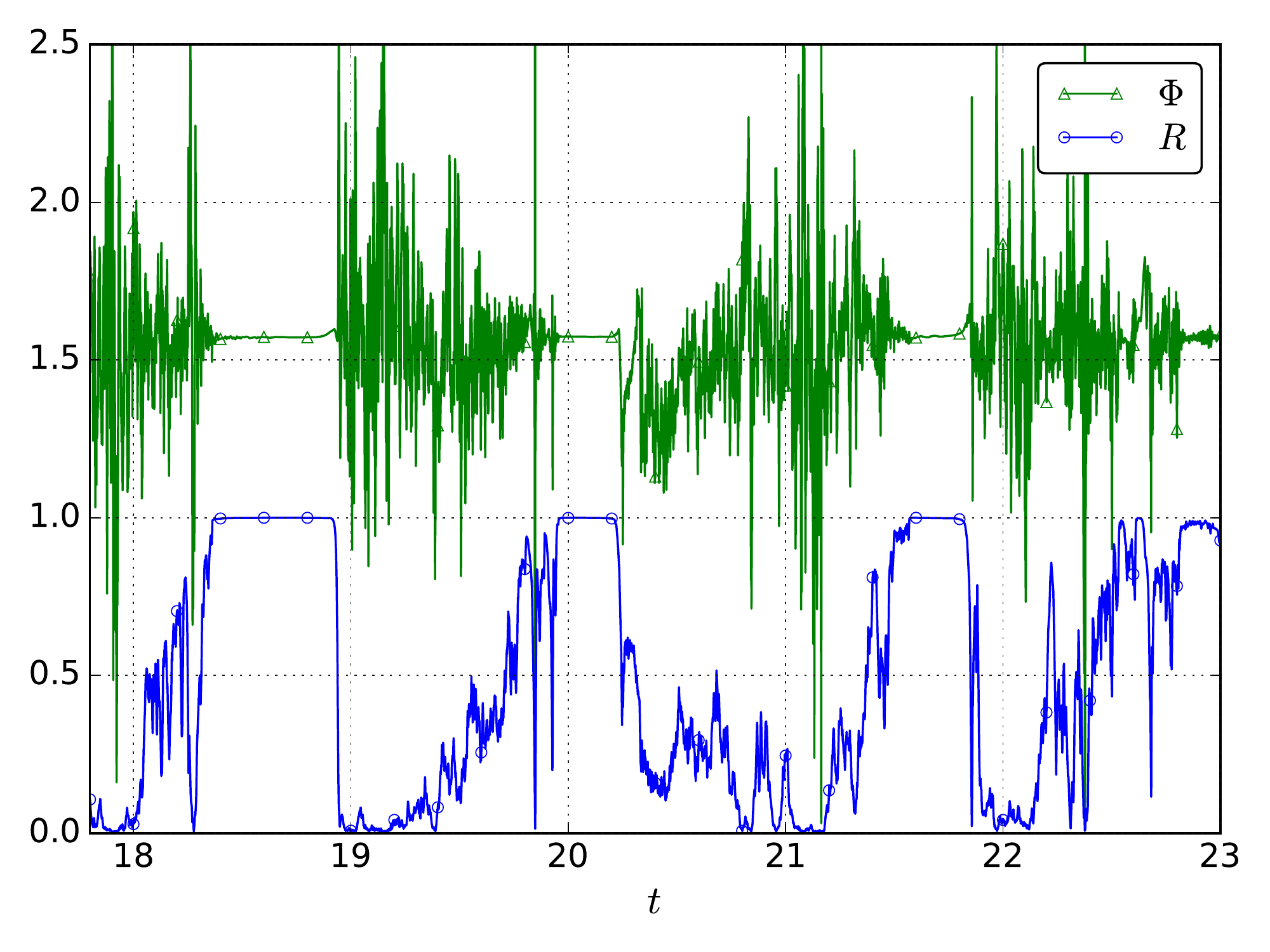}}
\vspace{-1mm}
\resizebox{0.50\textwidth}{!}{
\hspace{-0.2cm}\includegraphics{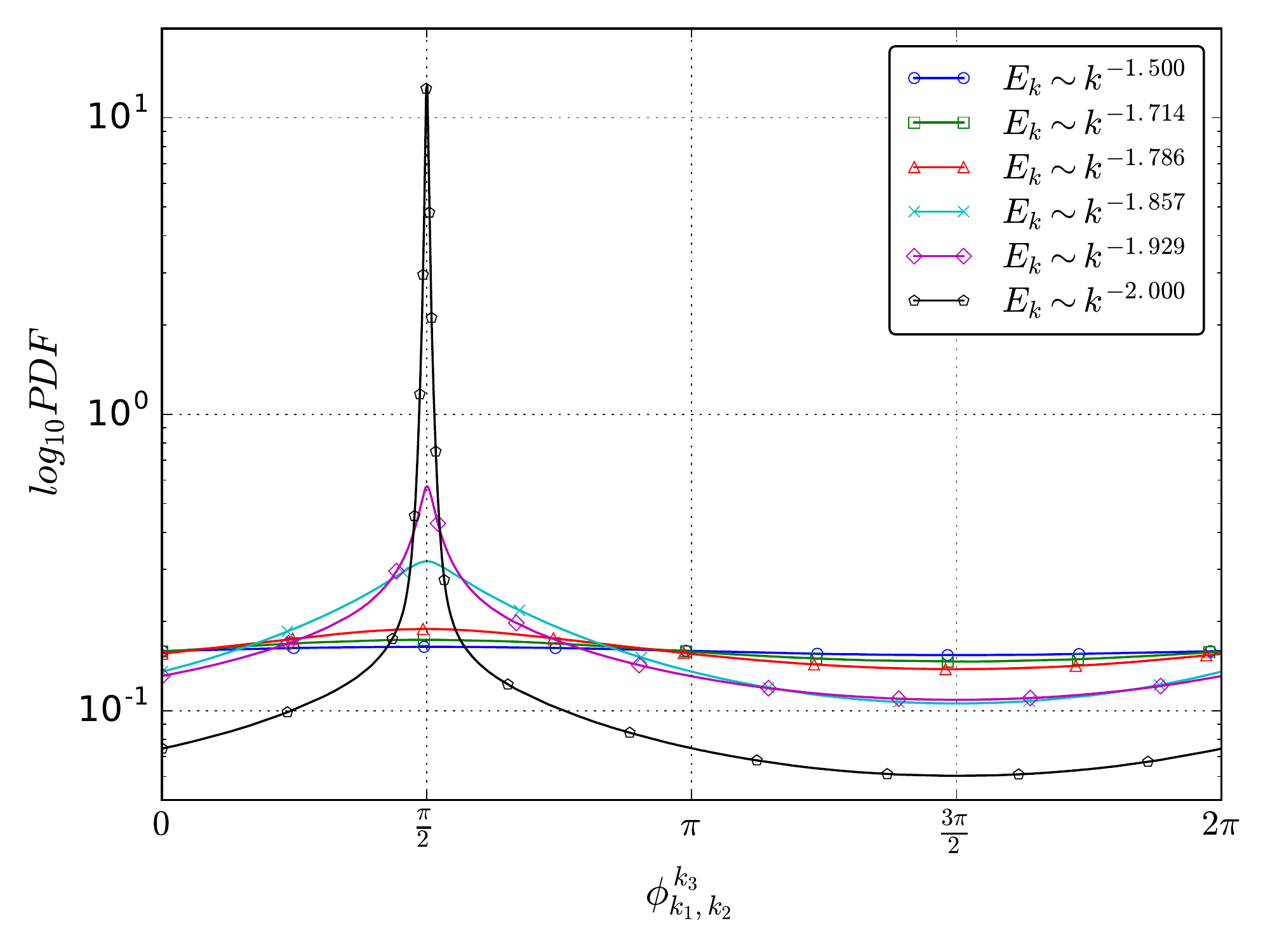}}
\resizebox{0.50\textwidth}{!}{
\hspace{-0.9cm}\includegraphics{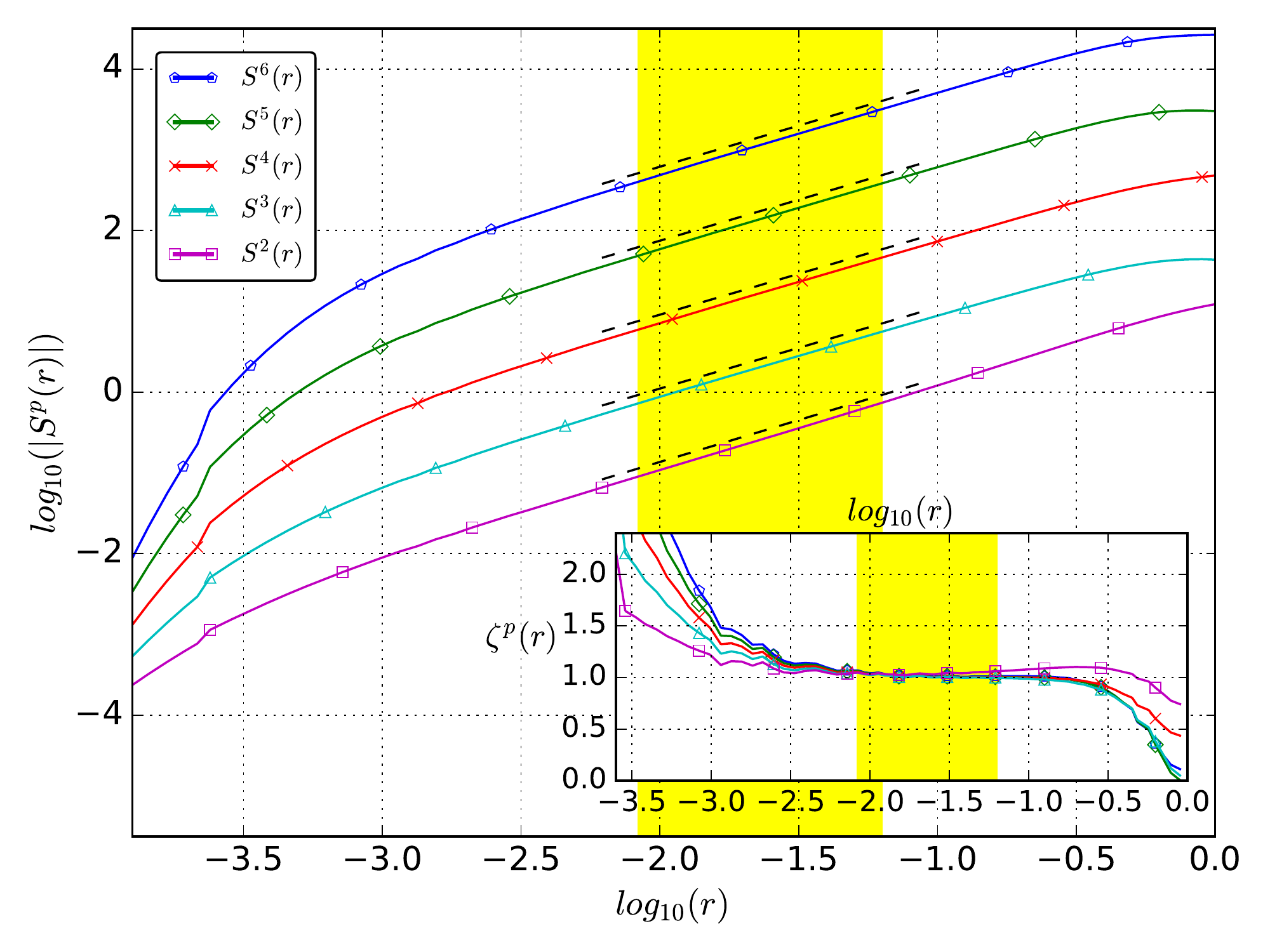}}
\resizebox{0.50\textwidth}{!}{
\hspace{-0.5cm}\includegraphics{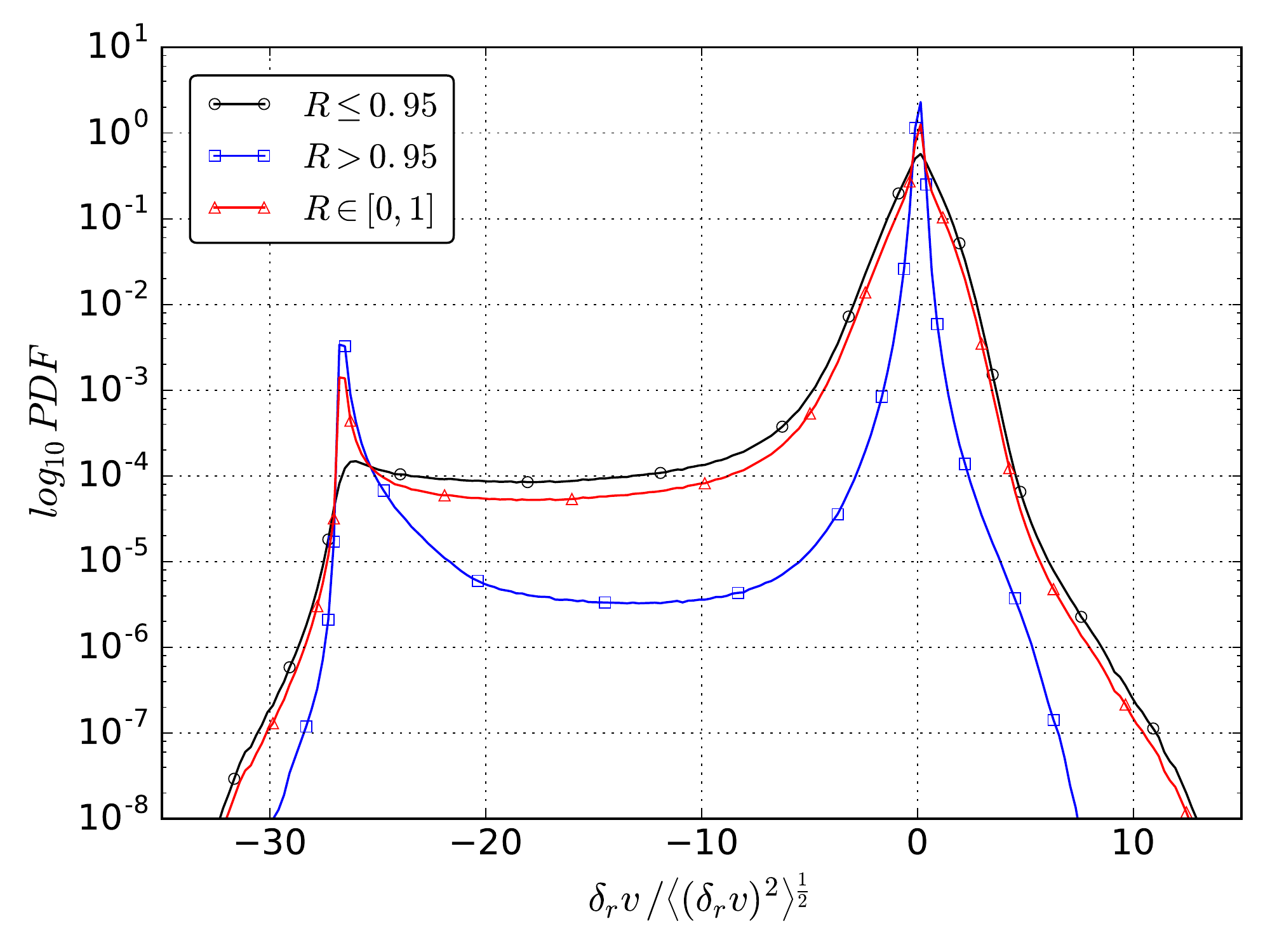}} 
\caption{Phase-Only Model data and statistics. \textbf{Top Left}: Time evolution of triad phase order parameter $R (t)$ (\textit{blue}) and $\Phi(t)$ (\textit{green}) for inertial-range triads. \textbf{Top Right}: Change in the PDF of triad phase as underlying spectrum slope is modified.  \textbf{Bottom Left}: Structure functions $S^{p}(r)$ as a function of $r$ and (inset) associated local slopes $\zeta_{p}(r)$. \textbf{Bottom Right}: Conditional time PDFs of  normalised velocity increments $\delta_{r} v \, / \left< (\delta_{r} v)^2\right>^{1/2}$ at inertial-range scale $k=450$. 
\label{fig:tseriesPO}
}
\end{figure}

Figure \ref{fig:tseriesPO} (top left panel) is a subset of time series of Kuramoto order parameters $R, \Phi$ for inertial-range triad phases evolving under phase-only system (\ref{eq:phas_only}). We see finite-duration intermittent synchronisation events, comparable to the full-PDE results (Figs.~\ref{fig:tseries} and \ref{fig:event_approach}). However the phase-only system shows some differences: the parameter $R$ can now take values close to $0$ and so $\Phi$ can vary more freely. Remarkably,  the system locks into the synchronised states on its own, without the influence of the energy dynamics. Figure \ref{fig:tseriesPO} (top right panel, black pentagons, $E_k \propto k^{-2.0}$) is the PDF of triad phases computed over time and over all inertial-range triads. This is similar to the triad-phase PDF obtained from the full PDE case, but the phase-only case has a more pronounced peak at $\pi/2$.

Figure \ref{fig:tseriesPO} (bottom panels) show structure-function local slopes and velocity increments PDF obtained from the phase-only model, to be compared directly with the full PDE results in figure \ref{fig:vel_incFP}. The well-resolved fat tails look qualitatively similar in both cases, leading to similar exponents $\zeta_p$ (see table 1 in the Ancillary Material for detailed fit values). In particular, the scaling $\zeta_p \approx 1$ is recovered in the case of the synchronised time periods ($R>0.95$) as seen in figure \ref{fig:tseriesPO}, bottom left panel. As for differences, the velocity-increment PDF shows more preference for positive values than in the full PDE case,  which is seen in physical space as small-amplitude oscillations with symmetric distribution about $\delta_r v = 0$. This difference is reduced when restricting to the synchronised periods ($R>0.95$) but the local symmetry with respect to $\delta_r v = 0$ is evident. This has a consequence of reducing the value of the exponent $\zeta_1^{\mathrm{abs}}$ in the phase-only case.

Based on previous work where inertial-range energy-spectrum exponent was found to vary with fractal decimation \citep{frisch2012,buzzicotti2016phase,michelehydro}, we study in figure \ref{fig:tseriesPO} (top right panel) the effect of the prescribed energy-spectrum inertial-range exponent on the phase-only model dynamics. We simply replace the fixed energy spectrum $E_k$ by $k^{2-\alpha} E_k$, where $\alpha$ ranges from $2$ to $1.5$, to mimic the variations observed. The triad phase PDF changes dramatically, to nearly total disorder for $\alpha=1.5$, in agreement with the results found in the full PDE simulation of fractal decimated Burgers \citep{buzzicotti2016phase}.

\subsection{Stability analysis of fully-synchronised state in the unforced case}

Consider a Galerkin truncation of system (\ref{eq:phas_only}) to resolution $N$: the relevant variables are the $N$ phases $\{\phi_k\}_{k=1}^N$, with $\phi_{-k} = -\phi_{k}$ due to the reality condition of the original scalar field $u(x,t)$. The system (\ref{eq:phas_only}), when unforced, has several fixed points (modulo a continuous symmetry, to be discussed below), given by $\phi_k = \pm \pi/2, \,k=1, \ldots, N$ (with $\phi_{-k} = -\phi_{k}$ always), where the notation $\pm$ means we can choose an arbitrary sign for the phase at each wavenumber $k$, and this choice somehow determines the stability of the fixed point. For example, the case $\phi_k =  \pi/2, \,k=1, \ldots, N$ is a fixed point corresponding to perfect alignment of triad phases, leading to the usual shock solution in 1D Burgers. As we will see below, this fixed point is stable in the case when the spectrum is of the form $(\sqrt{E_k} =) a_k \propto |k|^{-1}$, i.e. the classical shock spectrum. However, by varying the exponent in the above spectrum one can make this fixed point unstable. On the other hand, while keeping the classical shock spectrum it is possible to show that many of the other fixed points defined at the beginning of this paragraph are unstable.

It is important to keep in mind that in the unforced case the  physical system associated with (\ref{eq:phas_only}) has a continuous spatial translation symmetry $u(x,t) \to u(x+\epsilon,t)$, with $\epsilon$ arbitrary. This implies that the individual phases are not unique and one can in fact obtain the following induced symmetry:  $\phi_k(t) \to \phi_k(t) + \epsilon \,k,  \,k\in \mathbb{Z}.$ This is a symmetry of the evolution equations (\ref{eq:phas_only}), because the RHS of these equations depend on the triad phases $\phi_{k_1}+\phi_{k_2} - \phi_{k_1+k_2}$ and these are invariant under the induced symmetry. This symmetry has two consequences: 

First, being a continuous symmetry, it means there are only $N-1$ independent phases out of the $N$ phases $\{\phi_k\}_{k=1}^N$ ($\phi_{-k}$ being trivially dependent on $\phi_k$ via $\phi_{-k} = -\phi_{k}$). The ``lost'' degree of freedom corresponding to the symmetry leads thus to a neutral direction in the tangent space defined from the variables $\{\phi_k\}_{k=1}^N$. Explicitly this direction is parameterised by $\Delta \phi_k = k, \, k = 1, \ldots, N$.

Second, in the original set of variables $\{\phi_k\}_{k=1}^N$ any fixed point is in fact a continuous line of fixed points due to the symmetry. Thus, fixed points of system (\ref{eq:phas_only}) have unique coordinates in the reduced $(N-1)$-dimensional state space obtained after the symmetry has been ``factored out''. Instead of constructing an explicit change of coordinates to implement this reduction (using for example a set of $N-1$ independent triad phases), we take the approach of working in the original variables, while  keeping in mind the non-uniqueness of fixed points. To illustrate this situation, let us consider explicitly the fixed point $\phi_k =  \pi/2, \,k=1, \ldots, N$. Apply now the symmetry $\phi_k \to \phi_k + \epsilon \,k$ with $\epsilon = \pi$, so the fixed point is mapped to $\phi_k =  k\,\pi + \pi/2, \,k=1, \ldots, N$ or, restricted to the torus $[-\pi, \pi)^N$,
$\phi_k =  (-1)^k \pi/2, \,k=1, \ldots, N$. In words, if we flip the sign of the phases for all odd wavenumbers we obtain just another representative of the same fixed point. 

We can generalise this to the following statement: the set of fixed points of system (\ref{eq:phas_only}) can be represented by the following points:
$$\phi_k =  \frac{\pi\,s_k}{2}, \quad k=1, \ldots, N\,,$$
where $s_k = \pm 1$ and so the set $\{s_k\}_{k=1}^N$ represents a fixed point uniquely, provided we identify the set  $\{s_k\}_{k=1}^N$ with the set $\{(-1)^k s_k\}_{k=1}^N$. Using this identification, the number of different fixed points is $2^{N-1}$, a result that is equivalent to what one would obtain by simple counting in the reduced $(N-1)$-dimensional space of independent triad phases.

The motivation to introduce the above ideas is that the stability analysis of any fixed point of system (\ref{eq:phas_only}) will lead to a Jacobian matrix which possesses a null eigenvalue, with eigenvector corresponding exactly to the translation symmetry with neutral direction $\Delta \phi_k = k, \, k = 1, \ldots, N$.  Thus, relevant information about the stability of the dynamical degrees of freedom (i.e. those defined on the reduced $(N-1)$-dimensional state space) is obtained from the nonzero eigenvalues only. From these, eigenvalues with positive (negative) real part correspond to unstable (stable) directions whereas imaginary eigenvalues correspond to centres, always stable. 

Let us compute the equations determining the evolution of the system about the fixed point $\phi_k =  \pi/2, \,k=1, \ldots, N$. First, assume $\phi_k(t) =  \pi/2 + \delta\phi_k(t), \,k=1, \ldots, N$, where $\delta\phi_k(t)$ is a small perturbation. As always, $\phi_{-k} = -\phi_{k}$. Then, evaluate system (\ref{eq:phas_only}) in the unforced case, expanding the RHS to first order in the small perturbations. We obtain
\begin{eqnarray}
\hspace{-0.1cm}
\label{eq:perturb}
\frac{d}{dt}{\delta\phi}_k &=& \frac{k}{2} \, \sum_{k_1,k_2} \frac{a_{k_1} a_{k_2}}{a_k} \,\sin(\varphi_{k_1,k_2}^{k}) (\delta\phi_{k_1} + \delta\phi_{k_2} - \delta\phi_{k}) \, \delta_{k_1+k_2,k}  \,,
\end{eqnarray}
where $\varphi_{k_1,k_2}^{k}$ is evaluated at the fixed point: $\varphi_{k_1,k_2}^{k} = (\mathrm{sign}(k_1) +\mathrm{sign}(k_2) -\mathrm{sign}(k)) \pi/2\,. $ Now, write this system in standard form
$$\frac{d}{dt}{\delta\phi}_k  = \sum_{k'} J_{k k'} \delta\phi_{k'}\,,$$
in terms of the Jacobian matrix of time-independent components $J_{k k'}$ defined by
\begin{eqnarray*}
J_{k k'} &\equiv& \frac{\partial}{\partial (\delta\phi_{k'})}\left[ \frac{k}{2} \, \sum_{k_1,k_2} \frac{a_{k_1} a_{k_2}}{a_k} \,\sin(\varphi_{k_1,k_2}^{k}) (\delta\phi_{k_1} + \delta\phi_{k_2} - \delta\phi_{k}) \, \delta_{k_1+k_2,k}\right] \\
&=& \frac{k}{2} \, \sum_{k_1,k_2} \frac{a_{k_1} a_{k_2}}{a_k} \,\sin(\varphi_{k_1,k_2}^{k}) (\delta_{k_1 k'} + \delta_{k_2 k'} - \delta_{k k'}) \, \delta_{k_1+k_2,k}\\
&=& \frac{k}{2} \, \sum_{k_1} \frac{a_{k_1} a_{k-k_1}}{a_k} \,\sin(\varphi_{k_1,k-k_1}^{k}) (\delta_{k_1 k'} + \delta_{k-k_1, k'} - \delta_{k k'})\\
&=& \frac{k}{2} \left(  2 \frac{a_{k'} a_{k-k'}}{a_k} \,\sin(\varphi_{k',k-k'}^{k}) - \delta_{k k'} \sum_{k_1} \frac{a_{k_1} a_{k-k_1}}{a_k} \,\sin(\varphi_{k_1,k-k_1}^{k})  \right)\,.
\end{eqnarray*}

We now perform a detailed numerical study of the stability of system (\ref{eq:perturb}) in terms of the eigenvalues of the Jacobian matrix $J_{k k'}$, for a range of values of the spectrum exponent $\alpha$ in the formula $a_k \propto |k|^{-\alpha/2}$, and for a range of values of the resolution $N$. The case $\alpha=2$ corresponds to the usual shock solution. The case $\alpha=0$ corresponds to a hypothetical equipartition scenario. Values over the range $0\leq \alpha \leq 2$ are studied here. We recall that values in between were observed in our full PDE simulations of Fourier fractal-decimated 1D Burgers in a recent collaboration \citep{buzzicotti2016phase}, and also were studied in the previous subsection.  

\begin{figure}
\begin{center}
\resizebox{0.70\textwidth}{!}{
  \includegraphics{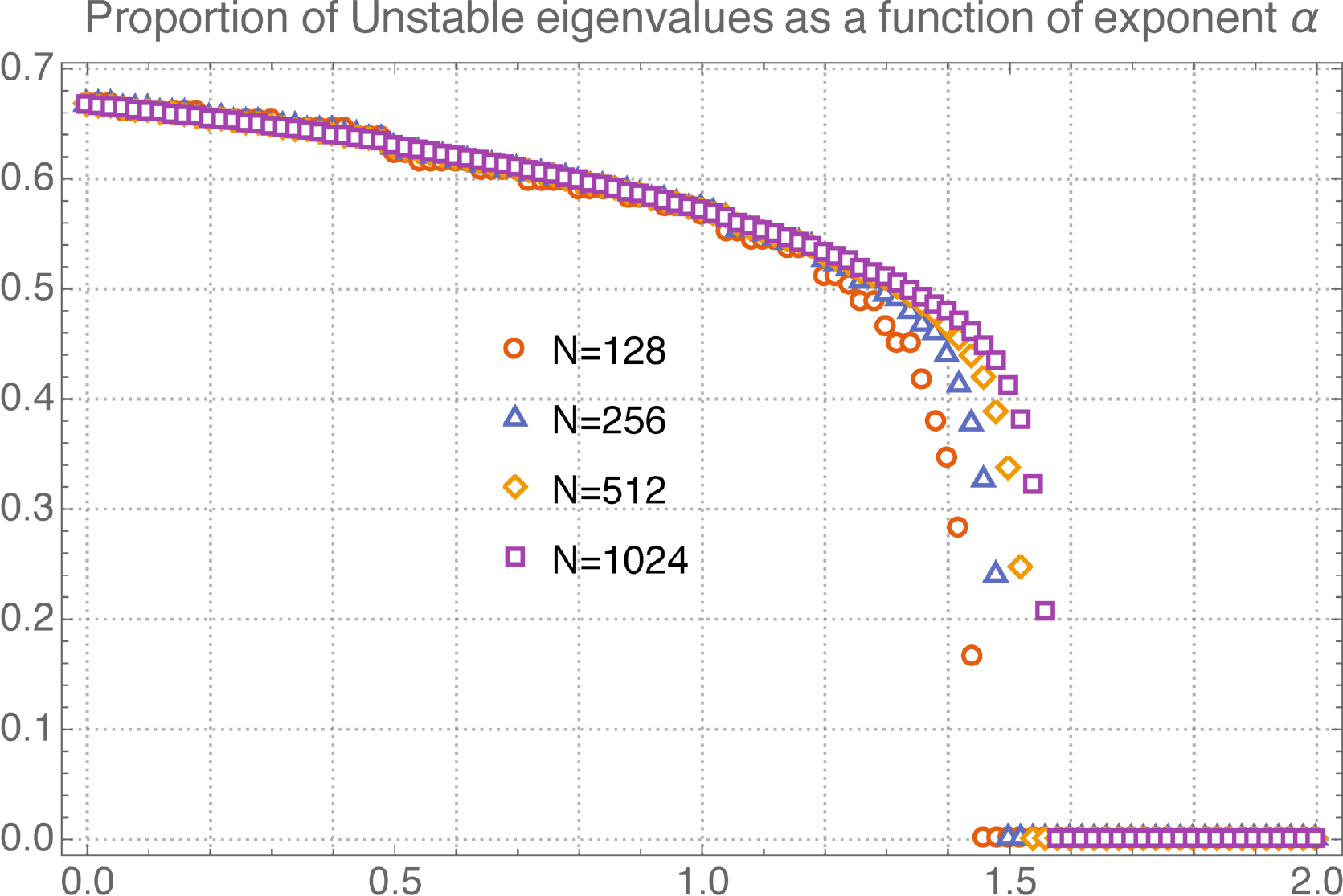}} 
\caption{Colour online. Stability study of fixed point $\phi_k =  \pi/2, \,k=1, \ldots, N$. For resolutions $N=128, 256, 512$ and $1024$ (red circles, blue triangles, orange rhombs and magenta squares respectively), plots of the proportion of unstable eigenvalues as a function of the spectrum exponent $\alpha$ in $a_k \propto |k|^{-\alpha/2}$. A sharp transition is observed at a value of $\alpha$ that depends on the resolution $N$.    
\label{fig:Stability_Study_All_Alpha_All_N}
}
\end{center}
\end{figure}

As explained above, for a given resolution $N$ we expect to observe generically 1 null eigenvalue and $N-1$ nonzero eigenvalues, with the null eigenvalue accounting for the spatial translation symmetry of the original system. A measure of the stability of the fixed point $\phi_k =  \pi/2, \,k=1, \ldots, N$ is provided by counting the number of unstable eigenvalues (i.e. eigenvalues with positive real part) as a proportion of the total number of non-zero eigenvalues $N-1$. Figure \ref{fig:Stability_Study_All_Alpha_All_N} shows the results of a numerical calculation of these eigenvalues, in four cases corresponding to the resolutions $N=128, 256, 512$ and $1024$. First we observe that, at any given resolution $N$, there is a sharp transition of this proportion, which goes from zero to nonzero abruptly as the parameter $\alpha$ is reduced continuously from the ``transition value'' $\alpha_{\mathrm{c}}(N)$ (`c' stands for `critical', borrowing notation from phase transitions in thermodynamics). This behaviour is due to a series of bifurcations where an eigenvalue with negative real part (i.e. a stable eigenvalue) becomes an eigenvalue with positive real part (i.e. unstable). Second, we observe that the proportion of unstable eigenvalues seems to converge, as $N$ grows, for any fixed value of the exponent $\alpha$ away from the transition values. Third, we observe that the transition value $\alpha_{\mathrm{c}}(N)$ moves towards $\alpha = 2$ as $N$ grows. Unfortunately, due to memory constraints it is difficult to go way beyond $N=1024$ in order to obtain a clear trend of the limit $N\to \infty$. The main practical difficulties are: (i) The Jacobian matrix dimensions grow as $N^2$. (ii) In order to obtain accurate results, the required numerical precision grows roughly as $N/2$. Notice that there are ways to mitigate precision-requirement issues by adding an exponential decay to the spectrum \citep{Agustin2017}, but with such a procedure the required precision still seems to be proportional to $N$, and the transition exponent is lowered with respect to the case of a pure-power-law spectrum, so more research is needed in order to establish whether adding an exponential decay to the spectrum can help in obtaining the asymptotic behaviour as $N\to\infty$.

\begin{figure}
\begin{center}
\resizebox{0.70\textwidth}{!}{
  \includegraphics{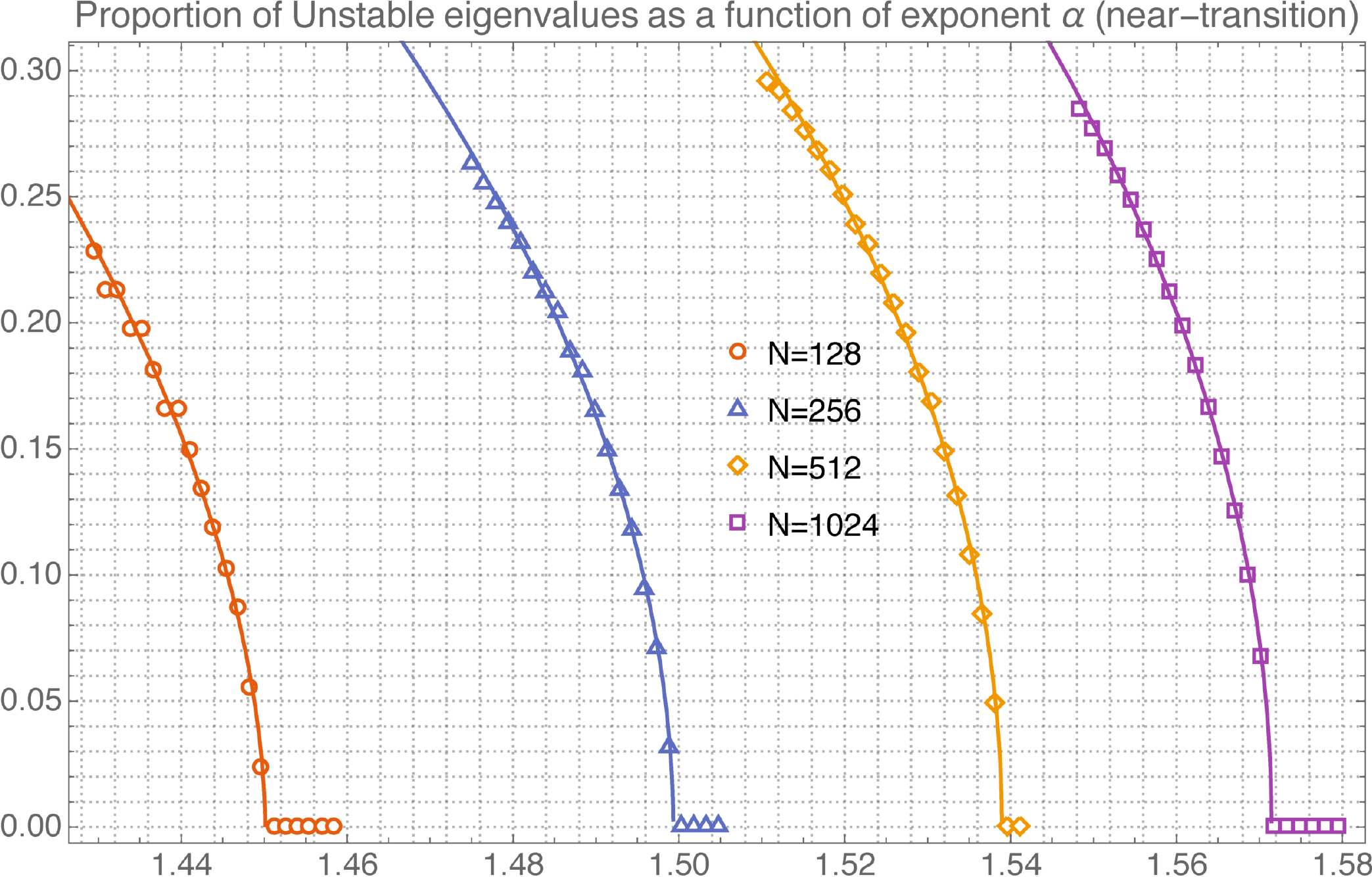}} 
\caption{Colour online. Stability study of fixed point $\phi_k =  \pi/2, \,k=1, \ldots, N$. Same parameters and colours as in figure \ref{fig:Stability_Study_All_Alpha_All_N}.  This is a more detailed study of the transition values of exponent $\alpha$ at each resolution. Solid lines represent the results of nonlinear fits done for each resolution separately, using all points that are shown in the figure. The fits are of the form $\mathrm{ratio} = C(N) (\alpha_c(N) - \alpha)^{z(N)}, \,\, \alpha \leq \alpha_c(N)$, where $C(N), \alpha_c(N)$ and $z(N)$ are fit parameters depending on the resolution $N$. 
\label{fig:Stability_Study_All_Alpha_All_N_Near_Critical}
}
\end{center}
\end{figure}

Notwithstanding the previous caveats, it is useful to look at the transition values of exponent $\alpha$ in more detail, if only to establish the classical behaviour expected in second-order phase transitions. Figure \ref{fig:Stability_Study_All_Alpha_All_N_Near_Critical} shows the result of this study. By considering more resolved ranges of $\alpha$ values close to the transition values, we repeat the calculation of the proportion of unstable eigenvalues, at each resolution. We use these data points to make a nonlinear fit of a hypothetical second-order phase transition of the form:
$$
\frac{\mathrm{\# \,\, unstable \,\, eigenvalues}}{N-1} = \left\{\begin{array}{lr}
C(N)  (\alpha_{\mathrm{c}}(N) - \alpha)^{z(N)}, &\alpha \leq \alpha_{\mathrm{c}}(N)\,,\\
0, & \alpha > \alpha_{\mathrm{c}}(N)\,,
\end{array}\right.
$$ 
where we introduce the resolution-dependent fit parameters $C(N), z(N)$ and $\alpha_{\mathrm{c}}(N)$. In practice we perform the fits on the natural logarithms of the data points and minimise the $L^2$ norm of the error. This procedure gives more weight to the data points that are close to the singularity. The results of the fit are shown in the following table:\\

\makebox[\textwidth][c]{
 \begin{tabular}{ c  c  c  c  c } 
 \toprule
 $N$ & $C(N)$  & $z(N)$& $\alpha_{\mathrm{c}}(N)$ & $R^2$\\ 
\midrule
 128 & 2.9 & 0.55 & 1.45011 &  0.99979\\ 

 256 & 2.6 & 0.52 & 1.49932   &0.99996 \\

 512 & 2.6  & 0.50  & 1.53895 & 0.99994\\

 1024 & 2.6  & 0.50 & 1.57147 & 0.99998\\
\bottomrule
\end{tabular}}
~\\

The last column is the coefficient of determination $R^2$ stemming from the nonlinear fit, which is an appropriate measure in this case due to the fact that we use natural logarithms of the data points, which do not vary too much. In the table, we rounded all entries to two significant digits, except for the $\alpha_{\mathrm{c}}(N)$ entries, which are used to obtain a more accurate estimate for the trend as $N \to \infty$. We notice first that the value of the prefactor $C(N)$ is quite stable. Second, the exponent $z(N)$ seems to be converging to the value $1/2$, the standard Landau value for the mean field theory of a second-order phase transition of a scalar field. Third, the transition values $\alpha_{\mathrm{c}}(N)$ do not seem to obviously converge, but a closer look on these (also supported by figure \ref{fig:Stability_Study_All_Alpha_All_N_Near_Critical}) shows that the distances between successive transition values decrease, at a rate that could be approaching a constant in the limit of $N$ large. In this plausible scenario, the criterion for convergence of the transition values $\alpha_{\mathrm{c}}(N)$ as $N\to \infty$ would be satisfied. The simplest model supporting a constant decrease rate for the distances between successive transition values is
$$\alpha_{\mathrm{c}}(N) = A - B N^{-C}, \quad A, B, C > 0$$ 
and fitting our $4$ data points from the previous table gives $A =1.7111, B =1.12188, C = 0.300571$, with a very good coefficient of determination $R^2 = 0.999999997$. Therefore, the sequence of transition values $\alpha_{\mathrm{c}}(N)$  would converge to $\alpha_{\mathrm{c}}(\infty) =  A = 1.7111.$ Although all this looks very good, one must take this result with care because it is not obvious that at resolution $N=1024$ an asymptotic regime has been reached.

\begin{figure}
\begin{center}
\resizebox{0.70\textwidth}{!}{
  \includegraphics{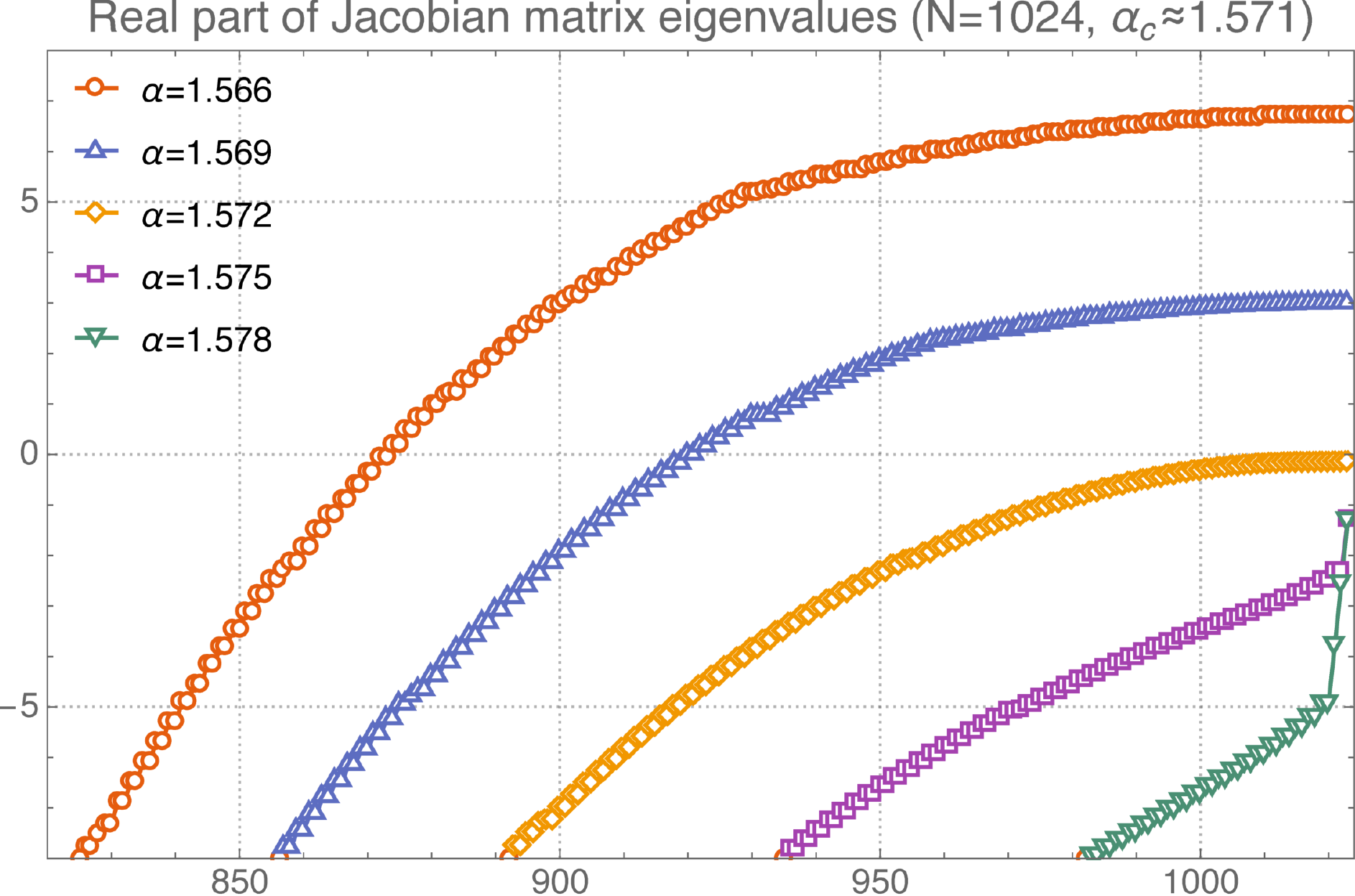}} 
\caption{Colour online. Plots, for several values of spectrum exponent $\alpha$ near the transition value $\alpha_{\mathrm{c}}$ (red circles, $\alpha=1.566$; blue triangles, $\alpha=1.569$; orange rhombs, $\alpha=1.572$; magenta squares, $\alpha=1.575$; green inverted triangles, $\alpha=1.578$), of the real part of the nonzero eigenvalues of Jacobian matrix with largest real part, centred about the eigenvalue zero for the purposes of better presentation. There are $1023$ eigenvalues and only the largest ones are shown. The lowest eigenvalues (not shown) take values as low as $-14000$.        
\label{fig:Eigenvalues_N_1024}
}
\end{center}
\end{figure}

For completeness, we show in figure \ref{fig:Eigenvalues_N_1024} the detailed distribution of real part of nonzero eigenvalues in the case $N=1024$, for several choices of exponent $\alpha$ near the transition value $\alpha_{\mathrm{c}}$. The plot shows only the eigenvalues with largest real part. It is evident that values of $\alpha$ less than the transition value $\alpha_{\mathrm{c}}$ develop eigenvalues with positive real part, with the largest eigenvalues distributed quite close to the maximum eigenvalue. In contrast,  values of $\alpha$ greater than the transition value $\alpha_{\mathrm{c}}$ produce negative eigenvalues only, with a quite sharp distribution of eigenvalues with respect to the maximum eigenvalue. 

So, to summarise, the study of stability of the fixed point $\phi_k = \pi/2, \,k =1, \ldots, N$ leads to a plausible mechanism to explain the sudden loss of synchronisation observed in the high-resolution simulations of the phase-only system as the spectrum exponent in $a_k \propto |k|^{-\alpha/2}$ is reduced from the classical-shock value $\alpha = 2$. Notice however that those simulations are performed under stochastic forcing (over large spatial scales), which provides extra time scales that are not taken into account in our simple stability analysis.

\section{Conclusion and Discussion}
\label{sec:conclusion}
We unveil a picture of turbulence in stochastically forced Burgers equation, whereby the Fourier \emph{triad phases} display a collective behaviour reminiscent of Kuramoto model \citep{Kuramoto84}. Intermittent triad phase synchronisation events give rise to strong and steady energy fluxes across scales and are seen in physical space as shock collisions. The fast time scale found for the onset of synchronisation suggests that the Fourier energy spectrum plays a passive role (via interaction coefficients): in fact, as a response to the triad-phase synchronisation the energy spectrum gains a dramatic coherence across the inertial range, while the triad phases maintain their synchronisation for a finite time interval before desynchronising again due to the stochastic forcing. We check the robustness of this picture by extending the range of forcing scales, reaching the same conclusions in terms of fast synchronisation and spectrum coherence, with the only difference that triad phase alignments are imperfect, a scenario consistent with the co-existence of many shocks in physical space. In this more generic scenario we establish again a clear connection between triad phase synchronisation events and energy dissipation bursts. We further support the picture of triad phase alignments by demonstrating the same synchronisation intermittency in phase-only simulations, i.e. with a prescribed energy spectrum. Velocity increments statistics are well reproduced by the phase-only system during synchronisation events, with structure-function exponents $\zeta_p \approx 1\quad (p>1)$. While we do not try to explain why the energy spectrum takes its characteristic profile $E_k \propto k^{-2}$ in the inertial range, we do produce a study of the \emph{effects} on the triad-phase dynamics of changing the exponent of a prescribed energy spectrum $E_k \propto k^{-\alpha}$ in the inertial range for the phase-only model. The results are striking: synchronisation events and triad phase alignment are drastically diminished as soon as $\alpha < 2$, with triad phases becoming totally disordered at $\alpha = 1.5$. This loss of synchronisation wipes out the fat tails in the velocity increments PDFs, so classical intermittency statistics are lost. These results provide a possible explanation of recent findings in fractal decimated Burgers equation \citep{buzzicotti2016phase,michelehydro}. To complement these results, we provide direct numerical evidence that the fully aligned state $(R=1, \Phi = \frac{\pi}{2})$ is a nonlinearly stable fixed point in the unforced phase-only model at resolution $N = 1024$, in the case $\alpha = 2$, and that it becomes unstable as $\alpha$ is lowered below some transition value $\alpha_{\mathrm{c}} \approx 1.571$, displaying the classical behaviour observed in second-order phase transitions. Looking forward, the next steps of this research will be to carry out similar analyses for 2D and 3D Navier-Stokes turbulence, to be published in a subsequent work.

As a final note motivated by the referees' comments, we discuss briefly the relation between our triad-phase synchronisation events and the evolution of complex singularities of the solution to Burgers equation. As we have shown in this paper, the synchronisation events of triad phases in Fourier space correspond to collision of shocks in physical space. Neglecting for the moment the fact that we have stochastic forcing, consider the solution to the \emph{unforced} small-viscosity Burgers equation defined on a periodic domain. In this solution, a single shock can be interpreted as a ``row of singularities'', which is an infinite set of isolated simple poles of the (now complex) function $u(z,t)$ in the complexified spatial variable $z=x + i y$, where $x$ is the usual spatial coordinate. This set is located along a line that is parallel to the imaginary axis: all poles have their real coordinate equal to $x_s$, the physical position of the shock, which may be a function of time. The imaginary coordinates of the poles are also functions of time. The positions of the poles satisfy a Calogero-type system of complex nonlinear ordinary differential equations  \citep{choodnovsky1977pole,calogero1978motion,bessis1984pole,bessis1990complex,senouf1996pole,senouf1997dynamics}. 
When viscosity is small, the pole that is closest to the real line remains at a distance to it that is proportional to the viscosity.

Thus, one may study the solution to the Burgers equations entirely from the point of view of the time evolution of the poles (as a finite truncation of the infinite-dimensional system of ordinary differential equations), or one may alternatively track the poles in post-processing using the numerical solutions obtained from direct numerical simulation of the partial differential equation 
\citep{morf1981analytic,sulem1983tracing,weideman2003computing,cichowlas2005evolution,caflisch2015complex}.

Regardless of the method used, the collision of two shocks can be interpreted as the result of the dynamical interaction between the complex poles located in two rows (one row corresponding to each shock). While such an event may be difficult to study from the dynamical systems' point of view in the complex plane $z = x+ i y$, it is still possible to analyse it numerically \citep{kida1986study}. The cited reference uses post-processing to look at the evolution of the poles that are closest to the real axis, for the solution with initial condition $u_0(x) = \frac{1}{2}(\sin(x) - \sin(2x))$. The process after the shocks form is quite interesting: the poles, while keeping a distance to the real axis, get closer and closer to each other as the physical shocks collide, which is the time when the poles merge. As the poles tend to merge, their distance to the real axis quickly halves: this is explained from the physical point of view of merging of two shocks of the same height (as is the case in this example), in that the shock height doubles after collision. The distance between the poles, in contrast, seems to decrease linearly with time until it reaches zero. In particular, as the shocks approach, the energy spectrum is filled with oscillations (as a function of wavenumber $k$) and a shortest period (in $k$ space) can be identified, that is proportional to the inverse of the distance between the shocks. So, as the shocks collide, the period of these oscillations tends to infinity, which means these oscillations disappear at the collision time. Of course, we observe this behaviour in our study in Section 2, and also to some extent in Section 3, although the presence of extra shocks away from the two colliding shocks implies that some oscillations still persist in the energy spectrum after the merging. What about the behaviour of the triad phases during these processes? In the case of two shocks on a collision course, with no other shocks present, one can show that right after the two shocks have completely formed, the triad phases $\varphi_{k_1 k_2}^{k_3}$ develop a characteristic hexagonal regular pattern spread across the $(k_1,k_3)$ wavenumber space, with basically a bi-modal distribution where about $3/4$ of the triad phases take values $\pi/2$, while $1/4$ of the triad phases take values $-\pi/2$, so that the order parameter $R$ is about $1/2$ (see for example the configuration near $t = 13.4$ shown in figure 2 of this paper, or the configuration at $t=16.282$ of a new simulation shown in the video described in figure 3 of the Ancillary Material: \url{www.youtube.com/watch?v=x0SokgDxmcU}). The pattern's typical size is proportional to the period of the spectrum's oscillations, i.e. it is inversely proportional to the distance between the complex poles. Thus, as the shocks get closer, the pattern maintains its regularity but is re-scaled further towards larger wavenumbers until the last patch of the pattern remains, which corresponds to the uni-modal distribution with all triad phases equal to $\pi/2$. This pattern's behaviour is obtained approximately if one assumes that the scalar field $u(z,t)$ is proportional to the sum of the two simple poles (along with their complex conjugates) located at the moving, ``colliding'' positions. The latter is a rough approximation to the exact solution given by the infinite sum over the simple poles along the two rows. Thus, the connection between pole-merging and triad-phase alignment is quite evident, at least descriptively, i.e. without regard to the dynamics.

Back to the more general setting of arbitrary initial conditions and general stochastic forcing, the behaviour of the system is not exactly as described above, for many reasons. First, the colliding poles are not always at the same distance to the real axis (i.e. the shock heights of the respective shocks are not equal), so the hexagonal pattern developed by the triad phases has a more complicated probability density function, rather than a simple bi-modal pattern. Second, the stochasting forcing may suddenly generate new rows of complex poles (i.e. new shocks in physical space). As is well known (see e.g. \cite{bessis1984pole}), when a new row of poles appears it does so from complex infinity instantaneously: more specifically, the distance from the leading pole to the real axis goes like $1/(t-t_0)$ where $t_0$ is the time at which the new row of poles appears. Thus, collision events are constantly being facilitated and modified by the stochastic forcing. Moreover, the presence of the stochastic forcing implies that the solution to the 1D Burgers equation is not given anymore by an infinite sum over simple poles; one would expect it to become a more general meromorphic function, still with an infinite set of simple poles so the idea of pole merging would still apply, but the detailed pole dynamics would be modified. Third and final, in the phase-only model that we propose the energy spectrum is kept fixed and equal to the time average of an actual 1D Burgers energy spectrum over a long simulation time. Thus, no information about the poles' relative distances is contained in the energy spectrum. However, the triad phases contain information about the position of the poles because shocks are formed when the triad phases align. Notice that shocks can form even in the unforced case because the fully-aligned state is nonlinearly stable, as we demonstrated. Thus, the phase-only dynamics \emph{with or without stochastic forcing} gives rise to new poles (or new arrangements into rows of poles) during the time evolution of the scalar field $u(z,t)$. In the stochastically forced case these poles seem to have a complicated dynamics unless the fully aligned state $R=1$ is attained (such a state is quite persistent but intermittent in the long run as can be seen in the video described in figure 4 of the Ancillary Material: \url{www.youtube.com/watch?v=OOY5Chd6SrE}). In subsequent work we will study the relationship between the dynamics of complex poles and the triad-phase alignment dynamics, in both full-PDE and phase-only models of 1D Burgers and other systems.

We acknowledge support from Science Foundation Ireland (research grant number 12/IP/1491) and COST-Action MP1305, and computational resources provided by the Irish Centre for High End Computing (ICHEC) under class C project \mbox{ndmat026c.} We thank D. Lucas, M. Buzzicotti, L. Biferale, M. Wilczek and J.-A. Arguedas-Leiva for enlightening discussions.

\bibliographystyle{jfm}
\bibliography{Burgers_Phase_Synchronisation}

\end{document}